\documentclass[traditabstract,twocolumn]{aa}

\usepackage{epsfig,graphicx,natbib}
\usepackage{longtable}
\usepackage{amssymb}
\def\Tcmb{\hbox{$T_\mathrm{CMB}$}}
\def\Trot{\hbox{$T_\mathrm{rot}$}}
\def\Tkin{\hbox{$T_\mathrm{kin}$}}

\def\nH2{\hbox{$n_\mathrm{H_2}$}}
\def\kms{\hbox{km\,s$^{-1}$}}
\def\PKS1830{\hbox{PKS\,1830$-$211}}
\def\cm-2{\hbox{cm$^{-2}$}}

\begin{document}

\title{An ALMA Early Science survey of molecular absorption lines toward PKS\,1830$-$211}
\subtitle{Analysis of the absorption profiles}

\author{S.~Muller \inst{1}
\and F.~Combes \inst{2}
\and M.~Gu\'elin \inst{3,4}
\and M.~G\'erin \inst{4}
\and S.~Aalto \inst{1}
\and A.~Beelen \inst{5}
\and J.\,H.~Black \inst{1}
\and S.\,J.~Curran \inst{6,7}
\and J.~Darling \inst{8}
\and Dinh-V-Trung \inst{9}
\and S.~Garc\'ia-Burillo \inst{10}
\and C.~Henkel \inst{11,12}
\and C.~Horellou \inst{1}
\and S.~Mart\'in \inst{3}
\and I.~Mart\'i-Vidal \inst{1}
\and K.\,M.~Menten \inst{11}
\and M.\,T.~Murphy \inst{13}
\and J.~Ott \inst{14}
\and T.~Wiklind \inst{15}
\and M.\,A.~Zwaan \inst{16}
}

\institute{
Department of Earth and Space Sciences, Chalmers University of Technology, Onsala Space Observatory, SE-43992 Onsala, Sweden
\and Observatoire de Paris, LERMA, CNRS, 61 Av. de l'Observatoire, 75014 Paris, France
\and Institut de Radioastronomie Millim\'etrique, 300, rue de la piscine, 38406 St Martin d'H\`eres, France
\and LRA/LERMA, CNRS UMR 8112, Observatoire de Paris \& Ecole Normale Sup\'erieure, Paris, France
\and Institut d'Astrophysique Spatiale, B\^at. 121, Universit\'e Paris-Sud, 91405 Orsay Cedex, France
\and Sydney Institute for Astronomy, School of Physics, The University of Sydney, NSW 2006, Australia
\and ARC Centre of Excellence for All-sky Astrophysics (CAASTRO)
\and Center for Astrophysics and Space Astronomy, Department of Astrophysical and Planetary Sciences, University of Colorado, 389 UCB, Boulder, CO 80309-0389, USA
\and Institute of Physics, Vietnam Academy of Science and Technology, 10 DaoTan, ThuLe, BaDinh, Hanoi, Vietnam
\and Observatorio Astron\'omico Nacional, Alfonso XII, 3, 28014 Madrid, Spain 
\and Max-Planck-Institut f\"ur Radioastonomie, Auf dem H\"ugel 69, D-53121 Bonn, Germany
\and Astron. Dept., King Abdulaziz University, P.O. Box 80203, Jeddah, Saudi Arabia
\and Centre for Astrophysics and Supercomputing, Swinburne University of Technology, Hawthorn, VIC 3122, Australia
\and National Radio Astronomy Observatory, P.O. Box O, 1003 Lopezville Road, Socorro, NM 87801, USA
\and Joint ALMA Observatory, Alonso de C\'ordova 3107, Vitacura, Santiago, Chile
\and European Southern Observatory, Karl-Schwarzschild-Str. 2, 85748 Garching b. M\"unchen, Germany
}

\date {Received  / Accepted}

\titlerunning{Strong absorption lines toward \PKS1830}
\authorrunning{Muller et al. 2014}

\abstract{We present the first results of an ALMA spectral survey of strong absorption lines for common interstellar species in the $z$=0.89 molecular absorber toward the lensed blazar \PKS1830. The dataset brings essential information on the structure and composition of the absorbing gas in the foreground galaxy. In particular, we find absorption over large velocity intervals ($\gtrsim$100\,\kms) toward both lensed images of the blazar. This suggests either that the galaxy inclination is intermediate and that we sample velocity gradients or streaming motions in the disk plane, that the molecular gas has a large vertical distribution or extraplanar components, or that the absorber is not a simple spiral galaxy but might be a merger system. The number of detected species is now reaching a total of 42 different species plus 14 different rare isotopologues toward the SW image, and 14 species toward the NE line-of-sight. The abundances of CH, H$_2$O, HCO$^+$, HCN, and NH$_3$ relative to H$_2$ are found to be comparable to those in the Galactic diffuse medium. Of all the lines detected so far toward \PKS1830, the ground-state line of ortho-water has the deepest absorption. We argue that ground-state lines of water have the best potential for detecting diffuse molecular gas in absorption at high redshift.
}
\keywords{quasars: absorption lines -- quasars: individual: \PKS1830\ -- galaxies: ISM -- galaxies: abundances -- ISM: molecules -- radio lines: galaxies}
\maketitle

\section{Introduction}


The spectroscopic study of absorption lines toward bright background continuum sources is a powerful technique for investigating the composition of the interstellar medium, as illustrated by the detection of the first interstellar molecules along the line-of-sight toward bright nearby stars (\citealt{swi37,mckel40,dou41}). Since the absorption signal is not diluted by distance, the sensitivity is only limited by the brightness of the background continuum source, allowing even rare molecular species to be detected. The discovery of molecular-rich absorption systems in four objects located at intermediate redshifts (0.24$<$$z$$<$0.89, e.g., see a review by \citealt{com08}) has opened up the possibility to explore the chemical contents of galaxies up to look-back times of half the age of the Universe. 

At intermediate to high redshifts, molecules can serve as interesting cosmological probes. For example, the evolution of the temperature of the cosmic microwave background (CMB) has been investigated up to $z$$\sim$3 using UV-band CO absorption-line systems observed in quasar spectra (\citealt{not11}). More recently, \cite{mul13} have obtained a precise and accurate measurement of the CMB temperature, \Tcmb=5.08$\pm$0.10\,K at $z$=0.89, based on a multi-transition excitation analysis of a set of different molecular species seen in absorption toward the blazar \PKS1830. This value is fully consistent with the value \Tcmb=2.725\,K$\times$(1+$z$)=5.14\,K at $z$=0.89, yielded by adiabatic expansion of the Universe.

Molecular absorbers are also widely used to probe the cosmological variations of fundamental constants of nature, such as the fine structure constant, $\alpha$, or the proton-to-electron mass ratio, $\mu$, (see e.g., \citealt{uza11}). Since different molecular transitions have different frequency dependence on the constants, the comparison of their velocity shift provides constraints on the constancy of the constants. The most used species are H$_2$, OH, CH$_3$OH, and NH$_3$. The tightest constraints to date yield null results down to $\Delta\mu$/$\mu$ of a few 10$^{-7}$ over 6--7\,Gyr (\citealt{kan11,bag13b}).

Moreover, molecular isotopologues provide a way to probe the nucleosynthesis enrichment of the Universe by comparing the isotopic ratios at different redshifts. Significant differences are found between $z$=0 and $z$=0.89, e.g., for $^{18}$O/$^{17}$O, $^{28}$Si/$^{29}$Si, and $^{32}$S/$^{34}$S (\citealt{mul06,mul11}). At a look-back time of $\sim$7\,Gyr \footnote{corresponding to $z$=0.89}, it is expected that low-mass stars contribute less than nowadays to the ISM pollution by their nucleosynthesis products, and the observed isotopic ratios at $z$=0.89 should mostly reflect the yields of massive stars.

Finally, from a ground-based observation point of view, the redshift can help shift lines (e.g., lines falling in bad atmospheric windows) to more favorable bands. This is the case, for example, of the fundamental 557\,GHz transition of ortho-water, which is completely impossible to observe from the ground at $z$=0, but could be detected in absorption at $z$=0.68 and $z$=0.89 by ground-based millimeter radio telescopes (\citealt{com97} and \citealt{men08}, respectively).


The most notable of the known redshifted molecular-rich absorbers is the $z$=0.89 galaxy toward the blazar \PKS1830\ (\citealt{wik96,wik98}). It has the highest redshift, the brightest background continuum, and the largest amount of absorbing material of all known molecular absorbers. The galaxy appears as a face-on spiral (\citealt{win02,koo05}) and acts as a gravitational lens, splitting the background blazar ($z$=2.5, \citealt{lid99}) into two main compact cores (NE and SW) separated by $\sim$1$''$ and embedded in a fainter pseudo-Einstein ring seen at cm-wavelengths (\citealt{jau91}). Molecular absorption features are seen along the lines-of-sight toward both compact lensed images (at $v$$\sim$0\,\kms\ and $v$$\sim$$-$147\,\kms, heliocentric frame and with $z$=0.88582, toward the SW and NE images, respectively), intercepting the disk of the $z$=0.89 galaxy on either side of its bulge. At millimeter wavelengths, the angular size of the continuum images corresponds to a scale on the order of one parsec at $z$=0.89, yielding a remarkably sharp pencil beam view through the disk of the absorbing galaxy.

The first unbiased radio spectral survey toward \PKS1830\ (at 30--50~GHz, \citealt{mul11}) has extended the molecular inventory up to a total of 34 different species toward the SW line-of-sight, making it the extragalactic object with the largest number of detected molecular species. The molecular abundances are found to be typical of Galactic diffuse-translucent clouds, and the physical conditions of the absorbing gas are well constrained (\citealt{hen08,hen09,mul13}). In contrast, only a handful of molecular species are detected toward the NE image, and the physical conditions of the absorbing gas in this line-of-sight remain poorly known.

Interestingly, \cite{mul11} discovered several additional velocity components, at $-$300, $-$224, $-$60, and +170~km\,s$^{-1}$ seen in the lines of HCO$^+$ and HCN. However, the continuum emission of the lensed blazar could not be resolved and the locations of these velocity components could not be determined.

The absorption line profiles are known to vary with a timescale of months (\citealt{mul08}), which is most likely due to morphological changes in the continuum emission from the blazar core/jet structures (\citealt{gar97,jin03,nai05}). So far, the variability has only been monitored in the lines of HCO$^+$ (and to a lesser extend of HCN, \citealt{mul08}) and CS (\citealt{sch14}). The time variations have the potential to reveal sub-parsec scale structures in the absorbing gas and the chemical correlation between different molecular species.


While the absorber in front of \PKS1830\ offers interesting opportunities as a cosmological probe, it is important to improve our knowledge of this source to understand and address systematics (e.g., \citealt{mul13} about the determination of the cosmic microwave background temperature at $z$=0.89 and \citealt{bag13b} about a constraint of the varation of $\mu$). The high angular resolution and sensitivity now available with the Atacama Large Millimeter/submilliter Array (ALMA) offer new possibilities in investigating the structure of the absorption and the molecular inventory along the two independent lines-of-sight toward \PKS1830. We have targeted the strongest absorption lines of some common interstellar molecules, such as CO, H$_2$O, CH, HCO$^+$, HCN, and NH$_3$, in addition to some other species which have transitions expected to be detectable within the same tuning bands. The main goals of this survey are:
\begin{enumerate}
\item resolve the structure of the absorption, locate the different velocity components, and obtain high signal-to-noise ratio line profiles;
\item investigate the chemistry and the nature of the absorbing gas. Is it similar to the Galactic diffuse component? Is there evidence of different (e.g., diffuse, dense) components? Or of different chemistry among the different velocity components?
\item study the time variations of the absorption profiles;
\item possibly detect new species in frequency ranges yet unexplored toward this source;
\item constrain the cosmological variations of fundamental constants and better understand the underlying systematics;
\item measure isotopic ratios from various isotopologues at a look-back time of more than half the present age of the Universe.
\end{enumerate}

Some of these goals (1 to 4) are addressed in this first paper, while others (5 and 6) will be discussed in forthcoming publications. The structure of the paper is as follows: in section \S\,\ref{sec:data}, we present the ALMA Cycle~0 observations and data reduction. In section \S\,\ref{sec:census}, we update the inventory of chemical species detected toward \PKS1830. The absorption profiles along the two lines-of-sight are analyzed in section \S\,\ref{sec:absprofiles}. Possible interpretations of the wide velocity spread seen along both lines-of-sight are reviewed in \S\,\ref{sec:thickmoldisk}. We investigate water as a tracer of molecular gas in absorption at high redshift in section \S\,\ref{sec:water}. A summary is given in section \S\,\ref{sec:conclusions}. The complete ALMA spectral scans are shown in Appendix\,\ref{appendix:spec}.

\section{Observations and data reduction} \label{sec:data}

The observations were undertaken during the ALMA Early Science Cycle~0 phase, between 2012 early April and mid-June (see Table\,\ref{tab:observations}). Four tunings were observed in spectral mode, at 100\,GHz (B3), 250\,GHz (B6), and 290 and 300\,GHz (B7) (see Table\,\ref{tab:center-freq}). For each tuning, four different 1.875\,GHz-wide spectral windows were set, each counting 3840 channels separated by 0.488\,MHz, providing a velocity resolution of 1--3\,\kms\ after Hanning smoothing.

In order to resolve the two lensed images of the blazar, we required the Cycle~0 extended configuration. This array configuration results in a synthesized beam of $\sim$$0.5''$ in B6 and B7, and $\sim$$2''$ in B3. Even though the latter beam exceeds the 1$''$ separation between the two images of the blazar, the high signal-to-noise ratio of the data allows us to resolve them in the Fourier plane (see below). The individual substructures of the blazar, i.e., the core/jet emission on milliarcsecond scale, remain unresolved. 

The data calibration was done within the CASA \footnote{http://casa.nrao.edu/} package following a standard procedure. Either Neptune or Titan were used for flux calibration and the absolute flux scale was set from a subset of short baselines for which their disks were not resolved. The flux scaling was then bootstrapped to all other sources and all baselines. We expect an absolute flux accuracy of $\sim$10\% in B3 and $\sim$20\% in B6 and B7.

\begin{figure}[h] \begin{center}
\includegraphics[width=8.8cm]{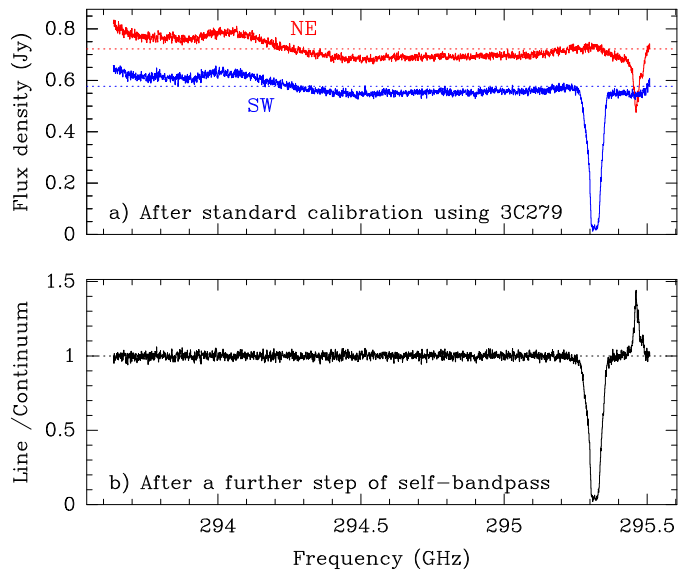}
\caption{Spectra of \PKS1830\ a) after a bandpass calibration using 3C\,279; b) after a further step of self-bandpass (i.e., the spectrum of the SW image has been divided by that of the NE image, and normalized to the continuum level).}
\label{fig:selfbp}
\end{center} \end{figure}

The bandpass response was calibrated from observations of the bright quasars J\,1924$-$292 or 3C\,279. We noticed that the spectral bandpass solution derived from 3C\,279 was not satisfactory, showing some residual ripples with amplitude of a few percent on the spectrum (see Fig.\,\ref{fig:selfbp}a). The origin of these ripples is probably related to imperfect linearity corrections used in the calculation of the spectral system temperature measurements, in addition to the fact that 3C\,279 had a high flux density of $\gtrsim$15\,Jy (B6 and B7) at the time of our observations. The incomplete corrections can introduce small frequency-dependent errors in measured amplitude and phase, resulting in a degraded spectral bandpass quality. By comparison, the spectral bandpass derived from J\,1924$-$292, of about 8\,Jy in B3, did not show any ripples. However, the bandpass of \PKS1830\ spectra could be considerably improved by a further step of ``self-bandpass'', using one image of \PKS1830\ to flatten the spectral bandpass of the other, as described in more detail below.

\begin{figure*}[ht!] \begin{center}
\includegraphics[height=\textwidth,angle=270]{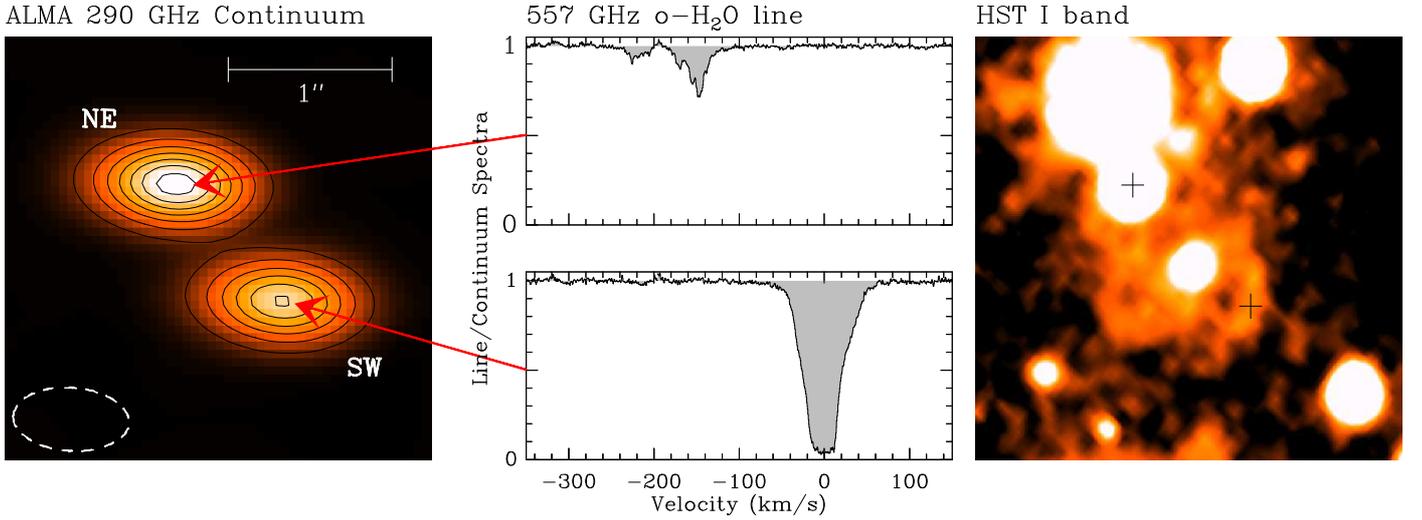}
\caption{Map of the 290\,GHz continuum emission of \PKS1830, showing the two resolved lensed images of the blazar ({\em left}, contour levels every 0.1\,Jy/beam, with the synthesized beam indicated as the dashed ellipse in the bottom left corner) and spectra of the fundamental 557\,GHz transition of ortho-water extracted toward each line-of-sight ({\em middle}, normalized to the continuum level of each lensed image). An I-band image obtained with the Hubble Space Telescope, revealing the foreground $z$=0.89 spiral galaxy, is also shown ({\em right}) with the same angular scale as the left panel. The two crosses indicate the position of the blazar images.}
\label{fig:clean}
\end{center} \end{figure*}

At mm/submm wavelengths, it is expected that the continuum emission of the blazar is reduced to the two compact images NE and SW, since the pseudo-Einstein ring has a much steeper spectral index ($\alpha$=1.5--2.0) than the two compact images ($\alpha$$\sim$0.7). For illustration, we show in Fig.\ref{fig:clean} the CLEAN-deconvolved map of the 290\,GHz continuum emission of \PKS1830, with spectra extracted toward each lensed image. This deconvolution step can be avoided, since the source can be simply modeled with two point-like components (with well known relative positions from VLBI observations) and the spectra can be directly extracted from the visibilities. This approach allows us to resolve the two images NE and SW, even at B3 in spite of the synthesized beam of $\sim$$2''$. We use the Python-based task {\sc uvmultifit} (\citealt{mar14}) to extract the spectra toward the NE and SW images from visibility-based analysis.

In a first step, the visibilities in the spectral channels with no absorption or atmospheric lines were fit with a model consisting of two point-like sources, where the relative positions were fixed and the flux densities left as free parameters. The visibility phases were then self-calibrated using this best-fit model, with one gain solution every 30\,seconds. The self-calibration improved the signal-to-noise ratio of the data typically by a factor of $\sim$2. Next, the self-calibrated visibilities were fit again with two point-like sources, but solved, on a spectral channel basis, for the flux density of the NE image and for the flux ratio SW/NE. The resulting ``spectrum'' contains the absorption lines toward the SW image, normalized to the flux density of the NE image, together with the absorption lines toward the NE image, but inverted, as seen in Fig.\,\ref{fig:selfbp}b. This ``self-bandpass'' strategy offers the advantages of removing the bandpass residuals from the bandpass calibrator, as discussed above, as well as from the atmospheric lines (dominated by O$_3$, see Table\,\ref{tab:O3atmline}). We noticed that the self-bandpass spectra occasionally showed a small residual spectral index, due to the intrinsic activity of the blazar and the time delay between the two lensed images. This was removed by a first order polynomial. Finally, the Doppler correction was applied in CASA, using the task {\em cvel} in the LSRK frame (Local Standard of Rest, kinematical definition), before the visibility fitting. The conversion to heliocentric frame was finally done as $v_{\rm hel}$=$v_{\rm LSRK}$+12.43 km\,s$^{-1}$, for easier comparison to previous studies.

\begin{figure}[h] \begin{center}
\includegraphics[width=8.8cm]{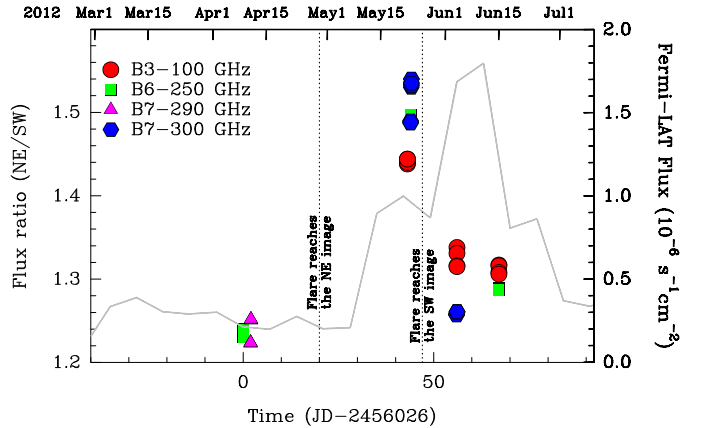}
\caption{Evolution of the flux ratio $\Re$ between the NE and SW images of \PKS1830 (marked symbols for each spectral window in our ALMA observations), overlaid over the $\gamma$-ray light curve (in grey) retrieved from the Fermi-LAT public archive. The dotted lines mark the time when the submillimeter flare begins to be seen in each image, according to the model by \cite{mar13}.}
\label{fig:contvar}
\end{center} \end{figure}

The ALMA Cycle~0 observations were taken serendipitously during a strong $\gamma$-ray flare (\citealt{cip12}). The flux-density ratio between the two lensed images of the blazar, $\Re$=$f_{NE}$/$f_{SW}$, shows a remarkable temporal and chromatic behavior (Fig.\,\ref{fig:contvar}). The time variations are due to the intrinsic activity of the background blazar and the geometrical time delay (25--27 days, \citealt{lov98,wik01,bar11}) between its two lensed images. More subtle is the chromatic variability, which was interpreted by \cite{mar13} as being caused by opacity effects in the jet base of the blazar (core-shift), modulated by the time delay. The sudden variations of $\Re$ and their coincidence with the $\gamma$-ray activity (an enhancement by a factor of seven) strongly suggest that they originate from the submillimeter counterpart of the $\gamma$-ray flare. The submm flux density of the blazar, on the other hand, does not show large variations ($<$10\%). As a result of the variability of $\Re$, the visibility sets observed at different epochs cannot be combined without introducing distortion in their interference pattern. Thus, we averaged the extracted spectra instead, after checking that the absorption profiles were not significantly affected by time variations (see \S\,\ref{sec:timvar}). The full resulting spectra are shown in Appendix\,\ref{appendix:spec}.

\begin{table*}[ht]
\caption{Journal of the observations.} \label{tab:observations}
\begin{center} \begin{tabular}{cccccccc}
\hline
Band          & Date of     & Bandpass   & Flux       & $N_{ant}$ $^{(a)}$ & $\delta$V $^{(b)}$ &  Noise $^{(c)}$ & Main targeted \\
              & observations & calibrator & calibrator &                 & (\kms)            & level (\%) & species \\
\hline
B3--100\,GHz & 2012 May 22 & J\,1924$-$292 & Neptune & 19 & 1.5 & 0.35 & HCO$^+$, HCN \\
             & 2012 Jun 04 & J\,1924$-$292 & Neptune & 17 & 1.5 & 0.39 &  \\
             & 2012 Jun 15 & J\,1924$-$292 & Neptune & 21 & 1.5 & 0.34 &  \\
\hline
B6--250\,GHz & 2012 Apr 09 & 3C\,279       & Titan   & 16 & 0.6 & 2.2 & CO\,(4-3), C\,I \\
             & 2012 Apr 09 & 3C\,279       & Titan   & 16 & 0.6 & 1.3 & \\
             & 2012 May 23 & 3C\,279       & Titan   & 16 & 0.6 & 0.9 & \\
             & 2012 Jun 15 & J\,1924$-$292 & Neptune & 20 & 0.6 & 0.7 & \\
\hline
B7--290\,GHz & 2012 Apr 11 & 3C\,279       & Titan   & 16 & 0.5 & 1.2 & H$_2$O, CH \\
             & 2012 Apr 11 & 3C\,279       & Titan   & 16 & 0.5 & 1.1 &  \\
\hline
B7--300\,GHz & 2012 May 23 & 3C\,279       & Titan   & 16 & 0.5 & 1.3 & NH$_3$, CO\,(5-4)\\
             & 2012 May 23 & J\,1924$-$292 & Neptune & 18 & 0.5 & 1.2 & \\
             & 2012 May 23 & J\,1924$-$292 & Neptune & 19 & 0.5 & 1.4 & \\
             & 2012 Jun 04 & J\,1924$-$292 & Neptune & 18 & 0.5 & 0.9 & \\
             & 2012 Jun 04 & J\,1924$-$292 & Neptune & 18 & 0.5 & 0.9 & \\
\hline
\end{tabular}
\tablefoot{(a) number of antennas in the array; (b) channel velocity spacing (half the effective velocity resolution, due to the adopted Hanning filtering); (c) noise level of the SW spectrum, normalized to the continuum intensity of the SW image.}
\end{center} \end{table*}

\begin{table}[ht] 
\caption{Central frequencies of the different spectral windows (SPW) of our survey. Each band is 1.875\,GHz wide.} \label{tab:center-freq} 
\begin{center} \begin{tabular}{ccccc} 
\hline 
Tuning & SPW\,1 & SPW\,2 & SPW\,3 & SPW\,4 \\ 
       & (GHz) & (GHz) & (GHz) & (GHz) \\
\hline 
B3--100\,GHz &  92.142& 94.017&104.142&106.017\\
B6--250\,GHz & 245.087&243.267&257.592&260.537\\
B7--290\,GHz & 282.573&284.448&294.572&296.446\\
B7--300\,GHz & 303.613&305.488&291.614&293.489\\
\hline 
\end{tabular} 
\end{center} \end{table}

\section{Inventory of species} \label{sec:census}

The inventory of species toward the SW image of \PKS1830\ has been largely expanded from the first unbiased spectral scan in the 7~mm band (30--50\,GHz, corresponding to 57--94\,GHz, in the absorber rest frame) performed with the Australia Telescope Compact Array (ATCA) by \cite{mul11}. The ALMA Cycle~0 observations add further species to the list, namely CH, H$_2$Cl$^+$, and NH$_2$, now giving a total of 42 different species (Table\,\ref{tab:inventory}). In addition, 14 different rare isotopologues have now been detected, the latest being $^{13}$CO, H$_2$$^{37}$Cl$^+$, and H$_2$$^{17}$O in the ALMA data. Toward the NE image, the number of detected species now reaches a total of 14, the ALMA Cycle~0 observations adding C, CH, CO, H$_2$O, H$_2$Cl$^+$, and NH$_3$.

The average spectral line density is of four lines per GHz in the 30--50\,GHz range covered by the ATCA 7\,mm survey. Although the ALMA Cycle~0 survey is heavily biased toward selected transitions of strong absorption lines, it covers a relatively large combined bandwidth of 4\,tunings$\times$4\,spectral windows$\times$1.875\,GHz=30\,GHz between the frequency limits, 91\,GHz and 307\,GHz. In this range, we detect a total of 27 transitions (not counting the multiple lines from hyperfine structure), that is, on average, one line per GHz. For the ALMA B3 tuning alone, the spectral line density reaches about two lines per GHz.

\begin{table*}[ht]
\caption{Census of species detected at $z$=0.89 toward \PKS1830. In bold face, first detections toward \PKS1830. Underlined, first extragalactic detections.} \label{tab:inventory}
\begin{center} \begin{tabular}{ll}
\hline
\multicolumn{2}{c}{\em Toward the SW image} \\
\hline
1 atom  & H $^{(d)}$, C $^{(in)}$ \\
2 atoms & {\bf CH} $^{(n)}$, OH $^{(d)}$, CO $^{(bcin)}$, {\bf $^{13}$CO} $^{(n)}$, CS $^{(afn)}$, C$^{34}$S $^{(f)}$, SiO $^{(jkmn)}$,  $^{29}$SiO $^{(km)}$, $^{30}$SiO $^{(m)}$, NS $^{(k)}$, SO $^{(kmn)}$, SO$^+$ $^{(k)}$ \\
3 atoms & \underline{\bf NH$_2$} $^{(n)}$, H$_2$O $^{(hn)}$, \underline{\bf H$_2$$^{17}$O} $^{(n)}$, C$_2$H $^{(ekmn)}$, HCN $^{(aefkmn)}$, H$^{13}$CN $^{(efkmn)}$, HC$^{15}$N $^{(fkmn)}$, HNC $^{(aefkm)}$, \\
& HN$^{13}$C $^{(efkmn)}$, H$^{15}$NC $^{(fkmn)}$, N$_2$H$^+$ $^{(ak)}$, HCO$^+$ $^{(aefkmn)}$, H$^{13}$CO$^+$ $^{(aefkmn)}$, HC$^{18}$O$^+$ $^{(fkm)}$, HC$^{17}$O$^+$ $^{(fkmn)}$, \\
& HCO $^{(kmn)}$, HOC$^+$ $^{(kmn)}$, H$_2$S $^{(f)}$, H$_2$$^{34}$S $^{(f)}$, \underline{\bf H$_2$Cl$^+$} $^{(n)}$, \underline{\bf H$_2$$^{37}$Cl$^+$} $^{(n)}$, HCS$^+$ $^{(m)}$, C$_2$S $^{(k)}$ \\
4 atoms & NH$_3$ $^{(ghn)}$, H$_2$CO $^{(cek)}$,  l-C$_3$H $^{(k)}$, HNCO $^{(km)}$, HOCO$^+$ $^{(m)}$, H$_2$CS $^{(k)}$\\
5 atoms & CH$_2$NH $^{(kmn)}$, c-C$_3$H$_2$ $^{(ekm)}$, l-C$_3$H$_2$ $^{(k)}$, H$_2$CCN $^{(k)}$, H$_2$CCO $^{(k)}$, C$_4$H $^{(k)}$, HC$_3$N $^{(ejkm)}$ \\
6 atoms & CH$_3$OH $^{(kln)}$, CH$_3$CN $^{(km)}$, NH$_2$CHO $^{(m)}$ \\
7 atoms & CH$_3$NH$_2$ $^{(km)}$, CH$_3$C$_2$H $^{(km)}$, CH$_3$CHO $^{(k)}$ \\
\hline
\multicolumn{2}{c}{\em Toward the NE image} \\
\hline
1 atom  & H $^{(d)}$, {\bf C} $^{(n)}$ \\
2 atoms & {\bf CH} $^{(n)}$, OH $^{(d)}$, {\bf CO} $^{(n)}$ \\
3 atoms & {\bf H$_2$O} $^{(n)}$, C$_2$H $^{(kn)}$, HCN $^{(fkmn)}$, HNC $^{(fk)}$, HCO$^+$ $^{(cfkmn)}$, \underline{\bf H$_2$Cl$^+$} $^{(n)}$\\
4 atoms & {\bf NH$_3$} $^{(n)}$, H$_2$CO $^{(k)}$  \\
5 atoms & c-C$_3$H$_2$ $^{(k)}$ \\
\hline \end{tabular}
\end{center} 
\vspace{1mm}
 \scriptsize{
{\it References:}
(a) \cite{wik96}; 
(b) \cite{ger97}; 
(c) \cite{wik98}; 
(d) \cite{che99}; 
(e) \cite{men99}; 
(f) \cite{mul06};
(g) \cite{hen08}; 
(h) \cite{men08}; 
(i) \cite{bot09}; 
(j) \cite{hen09}; 
(k) \cite{mul11};
(l) \cite{ell12,bag13a};
(m) \cite{mul13};
(n) this ALMA Cycle~0 survey.
}
\end{table*}

\section{Analysis of absorption line profiles} \label{sec:absprofiles}

In absorption, the depth $I_{abs}$, measured from the continuum level $I_{bg}$, is given by:
\begin{equation} \label{eq:tau}
I_{abs} = f_c I_{bg} (1-\exp{^{-\tau}}),
\end{equation}
\noindent where $\tau$ is the optical depth and $f_c$ is the fraction of background flux intercepted by the absorber (continuum covering factor). This depends on the frequency (chromatic structure of the continuum) and varies from species to species (chemical segregation in the absorber). The optical depth $\tau(v)$ is a function of the column density $N_{col}$ of the species, of the line excitation and of the line profile $\Phi(v)$:
\begin{equation} \label{eq:ncol}
\tau(v) = \frac{8\pi^3\mu^2S_{ul}N_{col}\Phi(v)}{3hQ(\Trot)} \frac{\left [ 1-\exp{ \left ( \frac{-h\nu}{k_B\Trot} \right ) } \right ]}{\exp{ \left( \frac{E_l}{k_B\Trot} \right ) }},
\end{equation}
\noindent where $h$ and $k_B$ are the Planck and Boltzmann constants, respectively, \Trot\ is the rotation temperature of the line, $Q(\Trot)$ the partition function, $E_l$ the energy of the lower level (with respect to ground state), $\mu$ the dipole moment, $\nu$ the line frequency, and $S_{ul}$ the line strength. The line profile $\Phi(v)$ is taken as a Gaussian. Introducing the coefficient $\alpha$ relating integrated opacity to column density, the column density is given by
\begin{equation} \label{eq:alpha}
N_{col} = \alpha \int \tau(v) dv.
\end{equation}
Under the physical conditions in the SW absorbing column (H$_2$ volume density \nH2$\sim$1000--2000\,cm$^{-3}$, and kinetic temperature \Tkin$\sim$80\,K, \citealt{hen08,hen09,mul13}), the excitation of highly polar species (i.e., all species in our survey, except CO and C\,I) is strongly radiatively coupled to CMB photons. Their rotation temperatures are \Trot$\sim$\Tcmb=5.14\,K, much lower than the kinetic temperature. In this case, $\alpha$ can be calculated from the parameters given in Eq.\ref{eq:ncol}. For the lines of CO and C\,I, the excitation is still subthermal, but collisional excitation starts to raise their rotation temperatures, so that \Tcmb$<$\Trot$<$\Tkin. We can use the non-LTE radiative transfer code RADEX (\citealt{vdtak07}) to estimate their $\alpha$ coefficients. Along the NE line-of-sight, the gas properties are poorly constrained but the absorbing column is likely even more diffuse than along the SW (\citealt{mul08}). Unlike the $\alpha$ coefficients of polar species, which remain constant as long as the density is moderate, the $\alpha$ coefficients of the CO and C\,I lines depend strongly on the physical conditions. Hereafter, we use the $\alpha$ coefficients given in Table\,\ref{tab:taudv} to derive column densities from integrated opacities.

\subsection{Velocity components} \label{sec:velocomp}

Besides the two previously known velocity components at $v$=0\,\kms\ (toward the SW image, heliocentric frame with $z$=0.88582, \citealt{wik96}) and $v$=$-$147\,\kms\ (toward the NE image, \citealt{wik98}), \cite{mul11} discovered three new velocity components, at $-$300\,\kms, $-$224\,\kms, and +170\,\kms\ seen in both the HCO$^+$ and HCN 1-0 lines in the ATCA spectra. Since the ATCA observations could not resolve the two lensed images of the blazar, these additional velocity components could not be assigned to any line-of-sight. 

\begin{figure}[h] \begin{center}
\includegraphics[width=8.8cm]{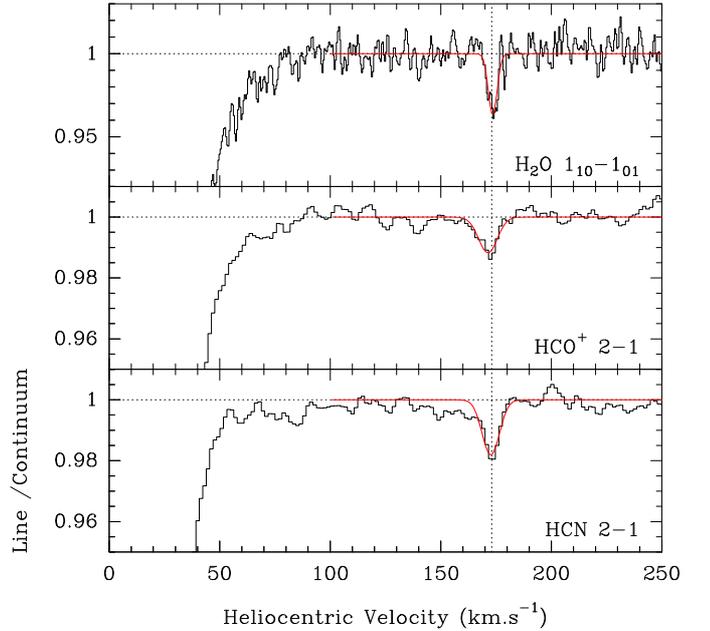}
\caption{Spectra of the H$_2$O, HCO$^+$, and HCN lines toward the SW image, showing the $v$=+170\,\kms\ velocity component besides the broad and deep feature at $v$$\sim$0\,\kms. The red curve shows a Gaussian fit of the line, with parameters given in Table\,\ref{tab:fit170}.}
\label{fig:v170}
\end{center} \end{figure}

Through the ALMA observations, the $v$=+170\,\kms\ component is now identified in the spectra of H$_2$O, HCO$^+$, and HCN toward the SW image (Fig.\,\ref{fig:v170}, and corresponding fit in Table\,\ref{tab:fit170}). Not only is the $v$=0\,\kms\ main SW absorption component very broad (with a FWZP of $\gtrsim$100\kms), but there is also gas shifted by +170\,\kms\ in velocity along the SW line-of-sight, which is quite remarkable considering that the galaxy is thought to be seen nearly face-on (see discussion in \S\,\ref{sec:thickmoldisk}). Assuming that the source covering factor is 100\% for this velocity component, we derive lower limits of column densities of $\sim$10$^{12}$\,\cm-2 and $\sim$2$\times$10$^{11}$\,\cm-2 for H$_2$O and HCO$^+$, respectively. Further assuming [H$_2$O]/[H$_2$]=5$\times$10$^{-8}$ and [HCO$^+$]/[H$_2$]=6$\times$10$^{-9}$ (their typical abundances in the Galactic diffuse component, see, e.g., \citealt{ger10,fla13}), we estimate a corresponding H$_2$ column density of $\sim$3--4$\times$10$^{19}$\,\cm-2. The H\,I absorption spectrum obtained with the Westerbork Synthesis Radio Telescope in October 1998, not resolving the two images and the pseudo-Einstein ring (i.e., integrated over the whole radio continuum emission), shows a continuous absorption covering the velocity interval $-$300\,\kms\ to +100\,\kms\ (\citealt{koo05}), i.e., the +170\,\kms\ velocity component has no apparent counterpart in H\,I \footnote{unless the H\,I absorption profile has changed since these observations. We are not aware of such report.}. The non-detection of H\,I absorption would put an upper limit of $\sim$10$^{19}$\,\cm-2 on the H\,I column density of this component, assuming a source covering factor of unity and a spin temperature of 100\,K. Thus, the $v$=+170\,\kms\ component is mostly molecular.

The $v$=$-$224\,\kms\ component is identified toward the NE image. In addition, the high sensitivity and spectral resolution of our ALMA observations allow us to detect multiple other velocity components, particularly well seen in the 557\,GHz rest frame water line and covering a velocity interval between $-$240\,\kms\ and $-$120\,\kms\ toward the NE image (Fig.\,\ref{fig:clean}, central panel). 

On the other hand, we do not detect any absorption feature corresponding to the $-$300\,\kms\ velocity component in either of the SW or NE ALMA spectra. This component was detected in both HCO$^+$ and HCN 1-0 lines in the ATCA data, although with a limited signal-to-noise ratio of $\sim$3\,$\sigma$. The ALMA spectra do not confirm this component and it remains unidentified. The strongest lines in the ALMA survey are the 557\,GHz water line (non-detection down to a rms of 0.9\% of the continuum level per 0.5\,\kms\ channel) and the HCO$^+$/HCN 2-1 lines (non-detection down to a rms of 0.2\% of the continuum level per 1.5\,\kms\ channel). Since the ATCA HCO$^+$ 1-0 absorption intensity of the $-$300\,\kms\ component reached 0.3\% of the {\em total} (NE+SW) continuum level and the 2-1 transition of HCO$^+$ should have a slightly higher opacity (assuming excitation by CMB photons as a minimum), we would expect to detect the line in the ALMA data. Its non-detection implies either that the signal in the ATCA spectra of both HCO$^+$ and HCN 1-0 lines was spurious, or that the absorption intensity of this component decreased since the ATCA observations.

\begin{table}[ht]
\caption{Results of Gaussian fittings for the $v$=+170\,\kms\ absorption feature toward the SW image.} \label{tab:fit170}
\begin{center} \begin{tabular}{lccc}
\hline
\multicolumn{1}{c}{Line} & $V_0$ $^\dagger$ & FWHM & Integrated opacity $^\ddagger$\\
    & (\kms)   & (\kms) &  (\kms) \\
\hline
H$_2$O $1_{10}$-$1_{01}$ & 173.5 (0.2) &  4.8 (0.6) &  0.187 (0.019)\\
HCO$^+$ 2-1            & 171.4 (0.5) & 10.0 (1.3) &  0.125 (0.013) \\
HCN 2-1                & 172.6 (0.4) &  9.3 (0.9) & 0.180 (0.015) \\
\hline
\end{tabular}
\tablefoot{$\dagger$ Heliocentric frame, taking $z$=0.88582; $\ddagger$ Assuming a continuum covering factor $f_c$=1.}
\end{center} \end{table}

\begin{figure*}[ht!] \begin{center}
\includegraphics[width=\textwidth]{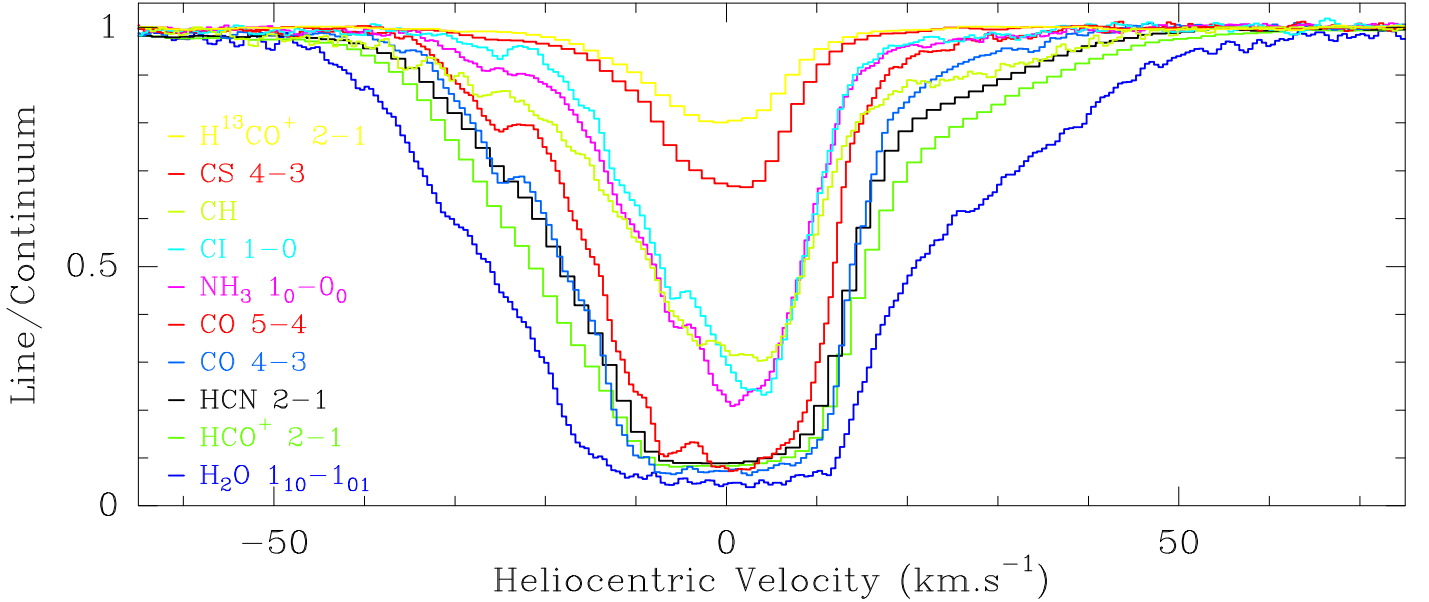}
\caption{Spectra of species with strong or saturated lines for the main $v$=0\,\kms\ velocity component toward the SW image. Lines are listed based on their increasing peak opacity, from H$^{13}$CO$^+$ to H$_2$O.}
\label{fig:strongline}
\end{center} \end{figure*}

\subsection{The absorption along the SW line-of-sight}

\subsubsection{Saturation and continuum source covering factor} \label{sec:saturation}

A few species in our survey -- namely H$_2$O, HCO$^+$, HCN, and CO --  show a very deep absorption toward the SW image (at $v$$\sim$0\,\kms). The lines (see Fig.\,\ref{fig:strongline}) are clearly saturated, as indicated by the detection of some of their rare isotopologues. However, the absorption depth does not reach the full intensity of the continuum emission and so we can conclude that the absorbing clouds do not entirely cover the SW continuum source. We measure the saturation level by averaging the absorption intensity for channels in the range $v$=$-$2\,\kms\ to +3\,\kms. Accordingly, the saturation level ranges between $\sim$95\% of the continuum intensity for the water line, and $\sim$91\% for the HCO$^+$/HCN (2-1) lines. This can be interpreted in two ways: On the one hand, if we assume that the saturation level provides a direct measurement of the source covering factor, $f_c$, then we conclude that the different species have slightly different covering factors (i.e., they have different distributions and are not co-spatial). On the other hand, the values 1/$f_c$ can be considered as the solid angle of the continuum emission divided by that of the absorbing clouds. Assuming that the physical size of the absorbing clouds is the same for all species (i.e., same source covering factor), the different values of 1/$f_c$ for the transitions of H$_2$O, CO (4-3), and HCO$^+$/HCN (2-1), only have a weak dependence on the frequency (between 100 and 300\,GHz), suggesting that the size of the SW image remains roughly constant at our ALMA frequencies. In contrast, past VLBI observations of \PKS1830\ at cm-wavelengths (e.g. \citealt{gui99}) show that the continuum size of the SW image decreases as $\nu^{-2}$, due to the scattering by interstellar plasma in the $z$=0.89 galaxy. The absence of such an effect here is consistent with a turnover of the continuum size -- frequency relationship at high (ALMA) frequencies, where the plasma becomes transparent.

In any case, our observations show that the size of the SW continuum emission is only slightly larger (by 5--10\%) than the size of the absorbing clouds at mm/submm wavelengths. This should naturally introduce time variability in the absorption line profiles whenever substantial changes in the continuum morphology occur. The continuum changes are potentially significantly magnified close to the lens caustics. Past observations have shown significant variations of the absorption intensity of the different velocity components (\citealt{mul08,mul11,mul13} and \citealt{sch14}) but no shifts in velocity, or only small ones. The small variations of $\sim$1\,\kms\ in the centroid of methanol lines reported by \cite{bag13b} might be due, as they note, to the simplifying assumption of only one Gaussian component in the fit of the profile. 

The $v$=+170\,\kms\ feature reaches an absorption depth of 3--4\% of the SW continuum level. Given that it does not appear to be saturated and that the covering factor of the $v$=0\,\kms\ component is $f_c$=0.95 suggests that both components are aligned (i.e., not on two adjacent lines-of-sight). The relative location of the $v$=+170\,\kms\ and $v$$\sim$0\,\kms\ features along the SW line-of-sight and their potential relationship cannot be elucidated at the moment.

\subsubsection{Line wings}

\begin{figure}[h] \begin{center}
\includegraphics[width=8.8cm]{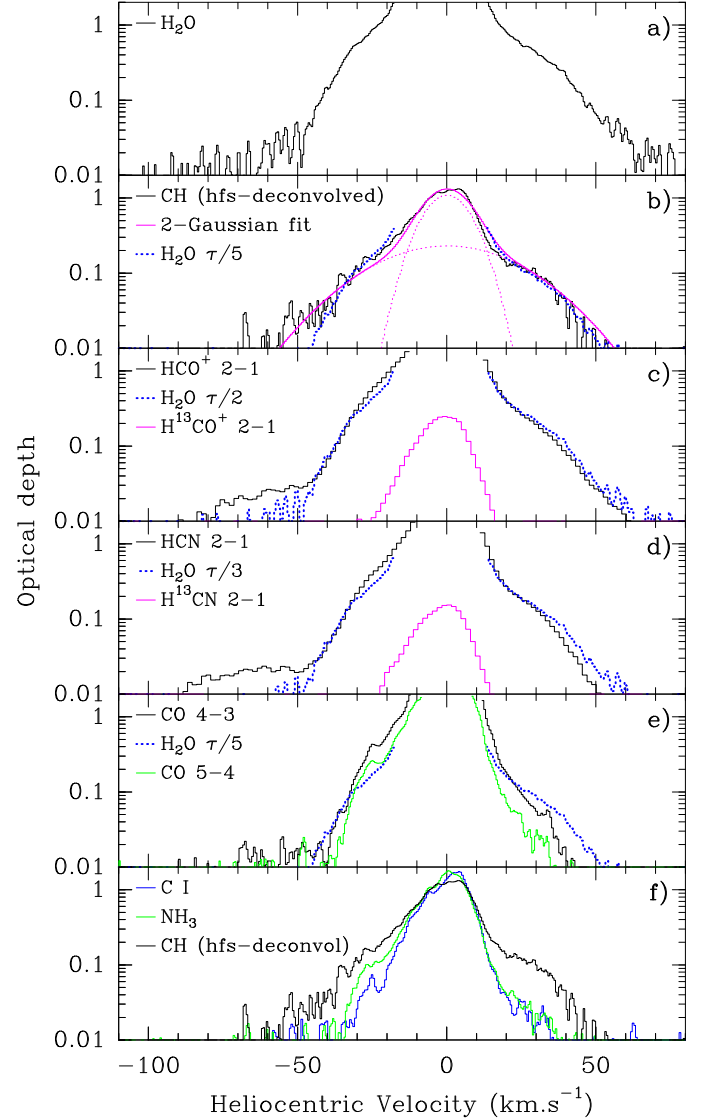}
\caption{Line wings of the $v$=0\,\kms\ velocity component toward the SW image. The ordinate denotes optical depth and is on logarithmic scale. Opacities are clipped at a level $\tau$=2, where they become less well determined due to uncertainties in the source covering factor.}
\label{fig:linewings}
\end{center} \end{figure}

The SW $v$$\sim$0\,\kms\ absorption covers a velocity spread exceeding 100~\kms\ (FWZP), evident from the water line (Fig.\,\ref{fig:strongline} and \ref{fig:linewings}a). This width has previously been noted by \cite{mul06} and \cite{men08} and remains puzzling considering the fact that the $z$=0.89 galaxy seems to be seen nearly face-on.

CH has been established as a robust tracer of H$_2$ in diffuse Galactic gas (\citealt{she08,qin10,ger10}). The CH absorption spectrum toward \PKS1830\,(SW) reaches an optical depth of $\tau$$\sim$1, and is not saturated. It is therefore a good species to investigate the absorption profile, after deconvolution from the line hyperfine structure (hfs). The hfs-deconvolved spectrum of CH can be decomposed into two components, one deep and narrow and the other broad and shallow (see Fig.\ref{fig:linewings}b). The narrow component corresponds to the absorption feature seen in all the non-saturated and weaker lines, e.g., H$^{13}$CO$^+$ 2-1 and CS 4-3 (Fig.\,\ref{fig:strongline}). The broad line wings are not due to a saturation effect (natural line broadening), since a simple scaling of the H$_2$O opacity reproduces remarkably well most of the profile of CH, HCO$^+$, and HCN (Fig.\ref{fig:linewings}), and the relative intensities of the three hyperfine components of CH are consistent with their expected ratios 0.83:0.33:0.17.

The fitting exercise, with two Gaussian CH components set at the same centroid velocity, results in a narrow component of FWHM=17\,\kms\ with an opacity of $\tau$$\sim$1.1, and a broad one of FWHM=53\,\kms\ with an opacity of $\tau$$\sim$0.23 (see Fig.\ref{fig:linewings}b). The corresponding integrated opacities are $\int\tau dv$$\sim$20 and 13\,\kms, respectively, i.e., in a ratio $\sim$60\% and 40\% of the total integrated opacity. Following Eq.\ref{eq:ncol} with \Trot=5.14\,K, we estimate CH column densities of 4.6 and 3.0 $\times$10$^{14}$\,\cm-2, respectively, for the narrow and broad component. Further assuming a fractional abundance [CH]/[H$_2$]=3.5$\times$10$^{-8}$ (\citealt{she08}), we derive a total (narrow+broad component) H$_2$ column density of 2.2$\times$10$^{22}$\,\cm-2\ along the SW line-of-sight, consistent with previous estimates in the literature (see \citealt{mul11} \S\,4.2 and references therein). We emphasize, however, that the narrow and broad components were not set apart in previous estimates based on molecular absorption.

For the velocity interval between $-$80 to $-$45\,\kms, we note a significant difference between the line wings of HCO$^+$/HCN and CH/H$_2$O (see Fig.\,\ref{fig:linewings}c and \ref{fig:linewings}d). The former show a small bump reaching apparent opacity of $\tau$$\sim$0.02, which is not present in the latter. This absorption feature has also been identified in the HCO$^+$ and HCN 1-0 spectra observed with ATCA by \cite{mul11}. It may originate from a molecular region illuminated by the background continuum at low frequency (i.e., for the HCO$^+$ 1-0 and 2-1 lines), but not a higher frequency.

\subsection{The absorption along the NE line-of-sight}

\begin{figure}[h] \begin{center}
\includegraphics[width=8.8cm]{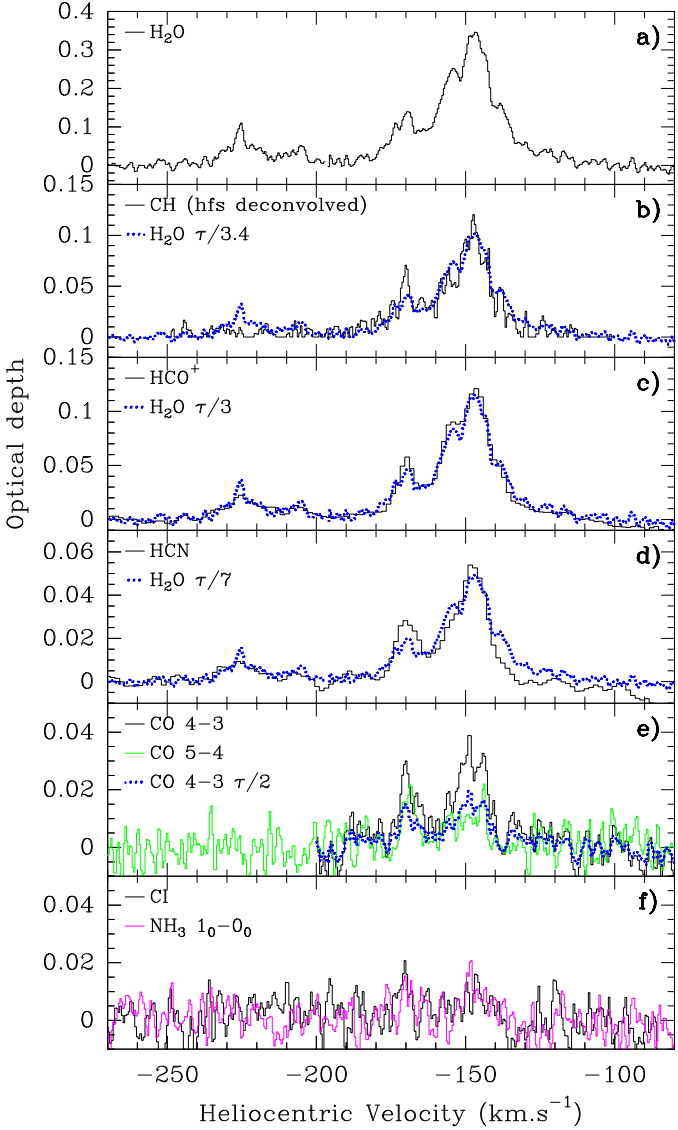}
\caption{Spectra toward the NE image (in optical depth scale, taking a source covering factor $f_c$=1 and thus providing lower limits to the actual opacities).}
\label{fig:specNE}
\end{center} \end{figure}

The absorption profile toward the NE image of \PKS1830\ is markedly different from the one toward the SW (Fig.\,\ref{fig:specNE}). Firstly, the absorption depths are much shallower and all lines, including the water lines, appear to be optically thin. Secondly, the absorption profile, as seen with ALMA, is decomposed in a series of narrow features, whereas the SW profile consists in a deep and broad absorption feature (not regarding the single weak +170\,\kms\ velocity component). This might be due to the much lower column density of molecular gas toward the NE image, which intercepts the disk of the foreground galaxy at a larger galactocentric radius of $\sim$4\,kpc, compared to $\sim$2\,kpc for the SW image. In contrast to the molecular absorption, the H\,I absorption is significantly deeper toward the NE image than toward the SW (\citealt{koo05}).

\begin{figure}[h] \begin{center}
\includegraphics[width=8cm]{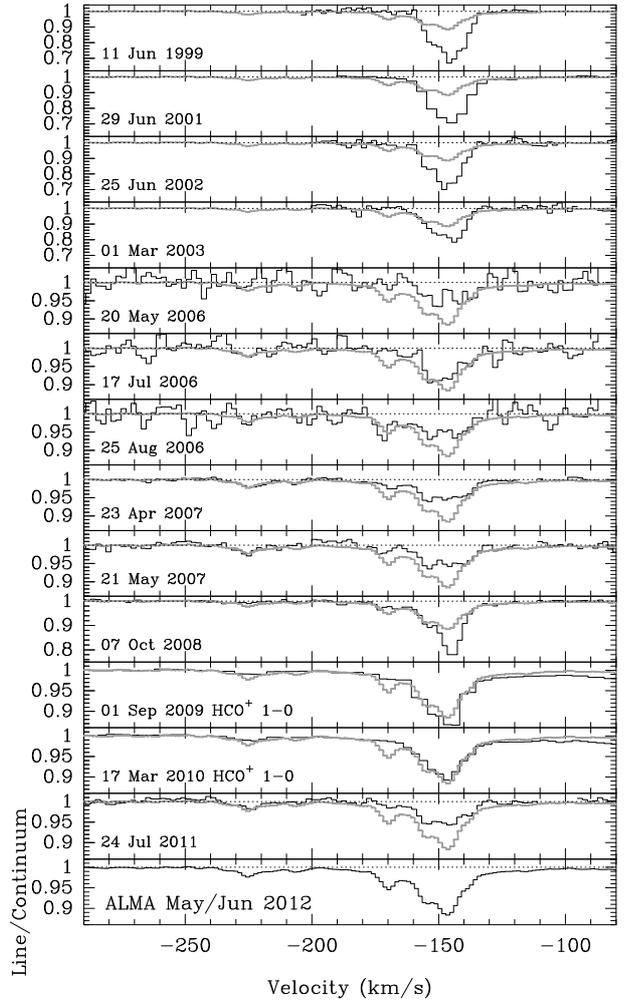}
\caption{Variations of the HCO$^+$ 2-1 (or 1-0 if specified) line for the NE absorption between 1999 and 2012 (spectra obtained by \citealt{mul08,mul11,mul13}). For direct comparison, the ALMA Cycle~0 spectrum is overlaid ({\em grey line}) on each box.}
\label{fig:varNE}
\end{center} \end{figure}

The physical conditions along the NE absorption have not yet been constrained due to the weakness of the absorption, but the time variability of the absorption profile suggests that the gas is more diffuse than toward the SW (\citealt{mul08}). The CO 4-3 and 5-4 lines detected in the ALMA Cycle~0 survey do not allow us to constrain further the physical conditions due to a strong degeneracy between the CO column density (or CO fractional abundance if we fix the H$_2$ column density) and the H$_2$ volume density.

We note that the surface area of the NE image is magnified by the factor $\Re$$\sim$1.3--1.5 with respect to that of the SW image, due to the gravitational lensing. We may then naturally expect a lower source covering factor than toward the SW image. Taking the absorption peak of the water line at face value, we obtain a strict lower limit of $f_c$$>$0.3. On the other hand, a close look at the NE absorption profile observed at different epochs (Fig.\,\ref{fig:varNE}) reveals independent variations of the different velocity components. For example, the spectrum taken on March 2010 appears identical to the spectrum taken by ALMA in June 2012, with the exception of the individual absorption feature at $v$=$-$170\,\kms. This indicates that the different narrow velocity components are not spatially aligned in the line-of-sight, and hence, the continuum covering factor (per velocity component) cannot be 100\%, giving 0.3$<$$f_c$$<$1.0. We note that the absorption profiles of the H$_2$O and HCO$^+$ lines, despite the factor of three in observed frequency, are remarkably similar (Fig.\,\ref{fig:specNE}b). Thus, the same absorbing regions should be illuminated at the two frequencies.

Within the whole velocity range $-$240 to $-$90\,\kms, we measure a total H$_2$O integrated opacity of 10\,\kms\ for the NE absorption. This value was obtained assuming a source covering factor of 100\%, and is therefore a lower limit. It corresponds to a total H$_2$O column density of $\sim$6$\times$10$^{13}$\,\cm-2\ and to a total H$_2$ column density of $\sim$10$^{21}$~\cm-2 (taking [H$_2$O]/[H$_2$]=5$\times$10$^{-8}$), that is, one order of magnitude lower than the column density toward the SW image.



\subsection{Relative abundances}

\begin{table*}[ht!]
\caption{Integrated opacities of strong absorption lines and species relative abundances.} \label{tab:taudv}
\begin{center} \begin{tabular}{cccccccc}
\hline
         &           &          & & \multicolumn{4}{c}{Integrated opacity $^a$ [Abundances relative to H$_2$]}  \\
Species  & $f_c$(SW) $^b$ & $f_c$(NE) $^b$ & $\alpha$ $^c$ &  SW blue wing $^d$ & SW line center $^e$ & SW red wing $^f$ & NE $^g$ \\ 
         &           &          & (\cm-2\,km$^{-1}$\,s)   & (\kms) & (\kms) & (\kms) & (\kms)  \\
\hline
CH   &    0.95 &    1.00   &    2.3E+13   &    2.38   [    3.5E-08     ]$^\diamond$ &   21.17    [   3.5E-08     ]$^\diamond$ &    2.07    [   3.5E-08     ]$^\diamond$ &    2.07    [   5.2E-08 ] \\
H$_2$O   &    0.95 &    1.00   &    6.3E+12       $^\dagger$ &   11.76   [    4.7E-08 ] &  sat.    [   -- ] &   10.27    [   4.8E-08 ] &    7.21    [   5.0E-08     ]$^\diamond$ \\
HCO$^+$   &    0.91 &    1.00   &    2.1E+12   &    7.08   [    9.5E-09 ] &  sat.    [   -- ] &    4.59    [   7.1E-09 ] &    2.55    [   5.9E-09 ] \\
HCN   &    0.91 &    1.00   &    3.5E+12   &    5.15   [    1.2E-08 ] &  sat.    [   -- ] &    2.86    [   7.4E-09 ] &    1.03    [   4.0E-09 ] \\
NH$_3$   &    0.95 &    1.00   &    1.2E+13      $^\ddagger$ &    1.22   [    9.3E-09 ] &   23.32    [   2.0E-08 ] &    0.51    [   4.5E-09 ] &    0.19    [   2.6E-09 ] \\
CO (4-3)   &    0.93 &    1.00   &    7.8E+15   &    4.37   [    2.2E-05 ] &  sat.    [   -- ] &    1.39    [   8.0E-06 ] &    0.65    [   5.5E-06 ] \\
CO (5-4)   &    0.95 &    1.00   &    1.5E+16   &    2.69   [    2.6E-05 ] &  sat.    [   -- ] &    0.57    [   6.2E-06 ] &    0.28    [   4.6E-06 ] \\
C\,I   &    0.93 &    1.00   &    7.6E+17   &    0.60   [    2.9E-04 ] &   22.32    [   1.2E-03 ] &    0.36    [   2.0E-04 ] &    0.22    [   1.8E-04 ] \\
\hline
\end{tabular}
\tablefoot{$a$: The uncertainties on the integrated opacities are of a few 0.01; 
$b$: source covering factor used to derive optical depth (see Eq.\ref{eq:tau}); 
$c$: see Eq.\ref{eq:alpha} for the definition of the $\alpha$ coefficients;
$d$: taken in the velocity interval [$-$40;$-$20]\,\kms;
$e$: taken between [$-$10;$+$10]\,\kms; 
$f$: taken between [$+$20;$+$50]\,\kms; 
$g$: taken between [$-$175;$-$135]\,\kms;
sat.: saturated ($\tau$$>$1);
$\diamond$ abundance fixed to scale that of other species in the same velocity interval;
$\dagger$ assuming an ortho-to-para ratio (OPR) of 3;
$\ddagger$ assuming OPR=1.}
\end{center} \end{table*}

In Table\,\ref{tab:taudv}, we list the opacities integrated over different velocity intervals for the strong lines of CH, H$_2$O, HCO$^+$, HCN, NH$_3$, CO, C\,I. From these integrated opacities and $\alpha$ coefficients (Eq.\ref{eq:alpha}), we can estimate column densities. Further, assuming that the abundance of CH or H$_2$O relative to H$_2$ are comparable to that in the Galactic diffuse medium (\citealt{ger10,qin10,fla13}), we derive the scaled relative abundances of all other species. We use CH as reference toward the SW absorption, since it is not saturated. Toward the NE, we use H$_2$O, which has the highest signal-to-noise ratio. We emphasize that since the lines in the NE absorption are optically thin, the relative abundances are not significantly affected by the unknown source covering factor.

A detailed analysis of the chemical composition (including other detected species) will be discussed elsewhere. Here, we note that the relative abundances in the wings of the SW absorption and in the NE absorption are within a factor of a few, and are also comparable to abundances in the Galactic diffuse interstellar medium. This confirms the results by \cite{mul11} concerning the fractional abundances of species detected in the ATCA 7\,mm survey.
The derived column densities and abundances of CO and C\,I are highly uncertain in the wings of the SW absorption and toward the NE image where the physical conditions are not constrained and the $\alpha$ coefficients may have different values. For all other species discussed here, the $\alpha$ coefficients do not vary significantly within a wide range of physical conditions. Their relative abundances are thus more robust.

\subsection{Time variations of the line profiles} \label{sec:timvar}

Time variations of the absorption profiles toward \PKS1830\ were reported by \cite{mul08} after (sparse) monitoring of the HCO$^+$ 2-1 line over a period of 12 years. Recently, variations for other species such as CH$_3$OH and CS have also been reported (\citealt{bag13b,sch14}). The origin of these long-term variations is thought to be due to changes in the continuum emission (\citealt{mul08}) possibly associated with a precessing jet (\citealt{nai05}), although evidence of a periodicity of the absorption changes remains to be seen.

Hereafter, we take advantage of the high quality of the ALMA spectra to investigate the potential effect of the $\gamma$-ray flare during the ALMA observations on the absorption profiles (short-term variations), and compare them to previous spectra obtained at other facilities (long-term variations).

\subsubsection{Short term variations, during the ALMA observations}

The submm counterpart of the $\gamma$-ray flare has occured in the SW image at some time between May 22 and June 04, as suggested by the inversion of the chromaticity of the flux ratio (Fig.\,\ref{fig:contvar}). Significant but small variations can be seen in the line wings of the strong absorption of saturated species (CO 4-3 and HCO$^+$ and HCN 2-1, see Fig.\ref{fig:timvar-23May09Apr}--\ref{fig:timvar-15Jun22May}). The relative variations are stronger at high frequency. This is consistent if the intensity of the flaring region gets lower with decreasing frequency (dilution). No significant variation is found for the NE absorption. It may be that the signal-to-noise ratio is not sufficient to detect this.

\subsubsection{Long term (yearly timescale)}

Here, we compare the ALMA B3-100\,GHz (2012) spectra with past ATCA-3\,mm data taken on 2011 July 27 (\citealt{mul13}). The two datasets have a wide frequency range in common (i.e., 91.2\,GHz to 95.0\,GHz, sky frequency), revealing the simultaneous variations of numerous species. Because the NE and SW images were not resolved by the ATCA observations, we have recomposed the ALMA spectra of the two lines-of-sight, assuming $\Re$=1.35, $f_c$(SW)=92\%, and $f_c$(NE)=100\%. The overlay of the ALMA 2012 and ATCA 2011 data is shown in Fig.\,\ref{fig:overlayATCA}. From their comparison, we can see that the depth of the SW absorption has significantly decreased between 2011 and 2012, while the NE absorption have become slightly stronger. More quantitatively, we compare the SW integrated opacities of various species in Fig.\ref{fig:timevar-comparo}. A factor of about two is found between the integrated opacities from 2011 and 2012. We note that the integrated opacity of the SiO J=4-3 line has dropped more than threefold, the largest change of all species observed in common in the ATCA and ALMA data.

\cite{mul13} noticed that the SW absorption depths in their July 2011 ATCA-7\,mm observations had increased by a factor of two compared to the previous observations in September 2009 (\citealt{mul11}). Retrospectively, it appears that the July 2011 observations have been carried on at a very favorable time for absorption studies toward the SW image, as illustrated by the detection of the rare isotopologues $^{29}$SiO and $^{30}$SiO, as well as of new species, NH$_2$CHO, HCS$^+$, and HOCO$^+$. By the time of our ALMA observations, the SW absorption depths have returned to their level in 2009. In contrast, the absorption toward the NE image became very weak in 2011, compared to 2009 and 2012 data. 

In short, the absorption depths toward both the NE and SW images can change by a factor of at least two within a timescale of a year and may not be as predictible as suggested by the long-term trends in monitoring so far (\citealt{mul08,sch14}). Clearly, a regular monitoring with a timescale on the order of one month would be desirable and could trigger dedicated studies (e.g., search of new species, high signal-to-noise-ratio observations) when the absorption depths increases significantly.

\section{Discussion}

\subsection{On the wide velocity spread of the absorption profiles} \label{sec:thickmoldisk}

Given that the two lines-of-sight toward \PKS1830\ sample gas on either side of the intervening galaxy's bulge (or lens center of mass) and that both show an absorption profile with a wide velocity spread of $\gtrsim$100~\kms\ indicates that either we are looking through two very special regions of the disk, or that the broad profiles are a common property of the molecular gas component throughout the disk. In the Milky Way, such broad absorption profiles are found only in the Galactic center region (e.g., toward Sgr\,B2 or Sgr\,A*, \citealt{mar12}).

Combining a low angular resolution H\,I spectrum (i.e., not resolving the two images of the blazar nor the pseudo-Einstein ring) with high angular resolution radio continuum maps of \PKS1830, \cite{koo05} modeled the H\,I kinematics using an azimuthally-symmetric gas distribution with a flat rotation curve. With this simple model, they constrain the inclination of the $z$=0.89 galaxy in a range $i$=17$^\circ$--32$^\circ$, i.e., finding that the galaxy is nearly face-on, consistent with the HST observations (\citealt{win02} and Fig.\,\ref{fig:clean}, right). Interestingly, their derived H\,I linewidth is $\sigma_{HI}$=39--48~\kms, which is high compared to local spiral galaxies (\citealt{koo05}). In addition, the results of their modeling point toward high H\,I depletion toward the galaxy center. \cite{koo05} suggested that the large H\,I linewidth could be due to feedback from active star formation in the galactic disk, resulting in high turbulent velocities.

In any case, the modeling of the H\,I data and the interpretation of the wide velocity spread of the molecular absorption are complicated by the limited information on the gas kinematics given by the sparse coverage of the background continuum. What could be the causes for the broad absorption profiles seen in the molecular gas? Possible interpretations include:
\begin{enumerate}

\item {\em The inclination of the disk:} While the modeling of the H\,I spectrum and the HST image suggest that the galaxy is nearly face-on, an intermediate inclination could be sufficient for the lines-of-sight to intercept a long column of gas through the disk and explore a significant part of the velocity gradient or streaming motions in the plane, especially in the center. 

\item {\em Extraplanar components:} The broad profiles might be due to a halo envelope, such as the warm diffuse component around the Milky Way, or to high/intermediate velocity clouds (\citealt{mul63}).
A molecular outflow or molecular streamers, such as those observed in M82 (\citealt{wal02}) can propel significant amounts of molecular gas into a galaxy halo. In the case of the $z$=0.89 galaxy in front of \PKS1830, we do not see clear symmetric evidence from receeding/approaching velocity components. However, the evidence could be hampered by the fact that we sample only two individual lines-of-sight.

\item {\em A thick molecular disk:} There has recently been mounting evidence of the existence of a thick disk of diffuse molecular gas seen in nearby galaxies. \cite{gar92} were first to detected extraplanar CO emission out to $>$1\,kpc above the disk of the edge-on galaxy NGC\,891. More recently, deep interferometric plus single-dish observations of M51 by \cite{pet13} suggest that up to 20\% of the total molecular mass could be extraplanar, with a typical scale height of $\sim$200\,pc. \cite{cal13} further present a study of the velocity dispersion of the atomic and molecular gas components in 12 nearby face-on galaxies, and find, surprisingly, a CO velocity dispersion similar to that of the H\,I layer. A thick disk might be the result of vertical turbulence, with gas ejected above the plane by star formation and the fountain effect.

\item {\em A merger system:} The foreground galaxy could be a merger system (i.e., similar to the Antennae) instead of a simple spiral, although there is no clear evidence of this in the HST image of \PKS1830.
\end{enumerate}

Note that broad H\,I and OH absorption profiles ($\sim$100--150\,\kms) are also found in the $z$=0.68 (also apparently nearly face-on) absorber toward B\,0218+357 (\citealt{car93,kan03}).

\subsection{Water as a tracer of molecular gas in absorption} \label{sec:water}

H$_2$O is a molecular species that is now detected in an increasing number of high-redshift galaxies. In their Herschel-ATLAS sample of lensed ultra-luminous starburst galaxies, \cite{omo13} report that water lines (such as 2$_{02}$-1$_{11}$ or 2$_{11}$-2$_{02}$, at 988\,GHz and 752\,GHz rest frame, respectively) can have intensities almost comparable to high-J CO lines, making them potentially a key tool to study the extreme conditions in the dense and warm medium of luminous high-$z$ galaxies. The aim of this section is to investigate the interest of water as a tracer of molecular gas {\em in absorption at high-$z$}, i.e., not only limited to trace the emitting warm dense component, but also tracing the diffuse gas component, which is only detectable in absorption.

\begin{figure*}[ht!] \begin{center}
\includegraphics[width=\textwidth]{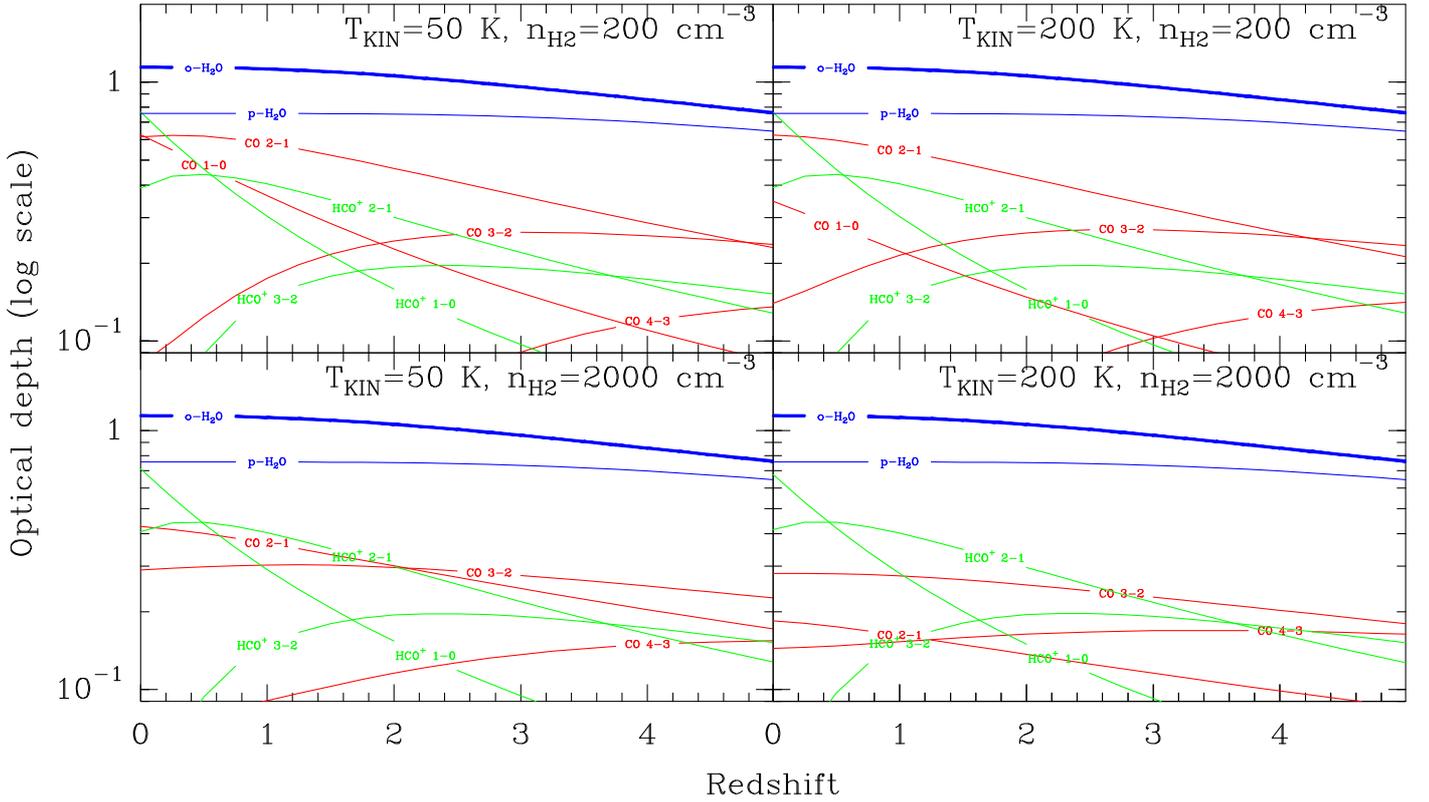}
\caption{RADEX-predicted opacities for transitions of CO ({\em red}), HCO$^+$ ({\em green}), and fundamental ground-state transitions of H$_2$O (ortho and para, {\em blue}), assuming a H$_2$ column density of 1.5$\times$10$^{20}$\,\cm-2, a line width of 1\,\kms, and fractional abundances [CO]/[H$_2$]=10$^{-5}$, [H$_2$O]/[H$_2$]=5$\times$10$^{-8}$, and [HCO$^+$]/[H$_2$]=6$\times$10$^{-9}$.}
\label{fig:opacity-predictions}
\end{center} \end{figure*}

In Fig.\,\ref{fig:opacity-predictions}, we plot the optical depths predicted with RADEX for several lines of CO, HCO$^+$, and H$_2$O, and several sets of H$_2$ density and kinetic temperature, as a function of redshift, taking \Tcmb$\propto$(1+$z$). We assume a H$_2$ column density of 1.5$\times$10$^{20}$\,\cm-2, a line width of 1\,\kms, and fractional abundances [CO]/[H$_2$]=10$^{-5}$ \footnote{In the diffuse molecular gas, the CO abundance strongly varies from $\sim$10$^{-7}$ to $\sim$10$^{-5}$ depending on the physical conditions (e.g., \citealt{she08})}, [H$_2$O]/[H$_2$]=5$\times$10$^{-8}$, and [HCO$^+$]/[H$_2$]=6$\times$10$^{-9}$, typical of abundances in the Galactic diffuse gas component (\citealt{she08,qin10,ger10,fla13}). We also assume collisions with only H$_2$. While the optical depth of CO lines varies largely, depending on the physical conditions of the gas \footnote{Due to its low electric dipole moment, the CO molecule is much easier to thermalize than molecules such as H$_2$O and HCO$^+$.}, we note that the optical depth of the first ground-state lines of ortho- and para-water remains roughly at the same level, and always higher than that of CO and HCO$^+$ lines. Because the excitation temperature of the H$_2$O lines is closely coupled to the CMB temperature, the H$_2$ column density can be directly estimated from observation of only a single water line, provided it is not saturated and the fractional abundance of water remains roughly constant, as is observed in the Galactic diffuse gas (\citealt{fla13}).

\begin{figure}[h] \begin{center}
\includegraphics[width=8cm]{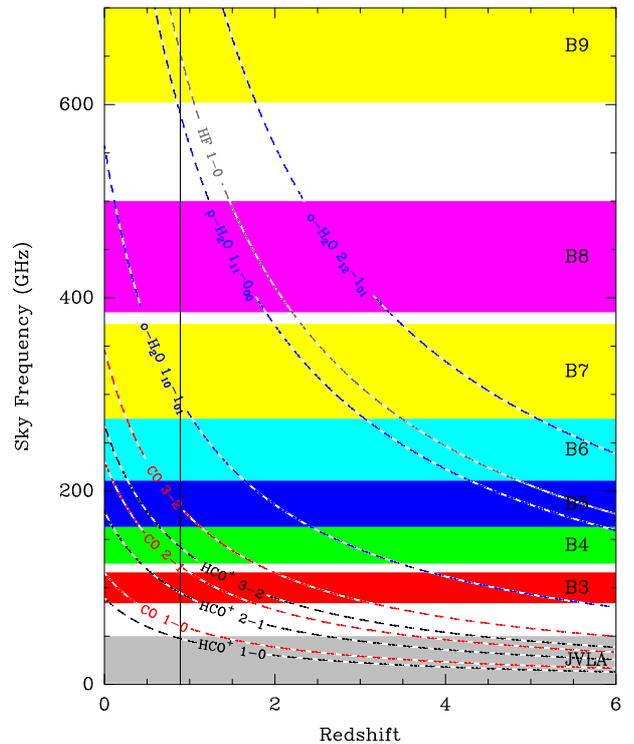}
\caption{Sky frequency of several tracers of molecular gas ({\em dashed blue:} H$_2$O; {\em dotted red:} CO; {\em dotted black:} HCO$^+$) as a function of redshift, with ALMA bands (labeled B3 to B9) and the Jansky Very Large Array (JVLA) frequency coverage overlaid.}
\label{fig:linevsz}
\end{center} \end{figure}

Fig.\,\ref{fig:linevsz} shows that the fundamental ground-state line of ortho-water (1$_{10}$-1$_{01}$, 557\,GHz rest frame) can be observed in ALMA bands for a wide redshift range, 0.1$\lesssim$$z$$\lesssim$5.6. For $z$$>$1.2, the ground-state line of para-water can also be observed, leading to the possibility of measuring the ortho-to-para ratio. As a result, water is an excellent tracer of the diffuse molecular gas component in absorption at high-$z$, observable from the ground (e.g., with ALMA). Number counts of submm galaxies indicate that a large pool of targets are available as background continuum sources for the sensitivity of ALMA (e.g., \citealt{cas14}).

According to Herschel observations, the HF 1-0 line may even have a stronger opacity than H$_2$O (\citealt{neu10,son10}). The HF molecule may thus be an alternative tracer for moderate redshifts. At very high redshifts, though, the elemental abundance of F may decrease, inhibiting the detection of HF (e.g., \citealt{lis11}).

\section{Summary and conclusions} \label{sec:conclusions}

We present the first results from an ALMA Early Science Cycle~0 spectral survey of the $z$=0.89 molecular absorber located in front of the blazar \PKS1830. Four spectral tunings at frequencies near 100, 250, 290, and 300\,GHz, were selected to cover strong absorption lines from common interstellar species, namely CO, H$_2$O, HCO$^+$, HCN, C\,I, and NH$_3$. The first results of this survey can be summarized as follows:

\begin{enumerate}

\item We enlarge the chemical inventory in this molecular-rich absorber with the detection of new species toward both lines-of-sight. In particular, the redshift of the absorber allows us to detect submillimeter lines observed recently with Herschel and inacessible or difficult to observe from the ground in the local $z$=0 ISM, such as those from CH, H$_2$O and H$_2$Cl$^+$. The inventory of species now reaches a total of 42 different species in the main absorption toward the SW image (plus 14 different rare isotopologues), and 14 species toward the NE image.

\item The observation of strong lines provide us with high signal-to-noise ratio spectra, with the two continuum images of \PKS1830\ spatially resolved. This allows us to reveal the absorption profiles along the two lines-of-sight with unprecedented detail. Toward the NE image, the absorption profile is resolved into a collection of narrow velocity components (5--10\,\kms\ wide) covering a wide velocity range of more than 100\,\kms. Toward the SW image, the main broad absorption also covers a velocity interval of more than 100\,\kms. In addition, the weak velocity component at +170\,\kms, previously detected but with unknown location, is now identified toward the SW image, i.e., presenting a remarkable large velocity offset of +170\,\kms\ with respect to the main ($v$=0\,\kms) absorption feature.

\item The large velocity interval seen along both lines-of-sight suggests either that the galaxy has an intermediate inclination and that we sample velocity gradients or streaming motions in the disk plane, that the gas in the foreground galaxy has a large vertical distribution (e.g., a thick molecular disk) or extraplanar components (e.g., high velocity clouds), or that the absorber has a more complex geometry than a simple rotating disk (e.g., it is a merger system).

\item We measure the continuum source covering factor toward the SW image from saturated lines, and find that it varies little with frequency, from $\sim$90\% at 100\,GHz for the HCO$^+$/HCN 2-1 line to 95\% at 300\,GHz for the H$_2$O 1$_{10}$-1$_{01}$ line.
Either the covering factor is different for the different species (i.e., they are not co-spatial), or the size of the continuum emission is roughly the same between 100 and 300\,GHz.

\item The ALMA observations were taken (serendipitously) at the time of a strong $\gamma$-ray flare of the background blazar. A study of the temporal and chromatic variations of the flux ratio between the lensed images of the blazar during the flare is reported by \cite{mar13}. Over the time span of the observations ($\sim$two months), we find only minor variations in the spectral line profiles, mostly in the wings of saturated lines.

\item Of all the lines detected so far toward \PKS1830, the H$_2$O 1$_{10}$-1$_{01}$ line (557\,GHz rest frame) has the deepest absorption. We argue that ground-state water lines are excellent probes of molecular absorption at high-redshift.

\end{enumerate}

An accompanying paper (\citealt{mul14}) focusses on the first extragalactic detection of the chloronium ion, H$_2$Cl$^+$, toward \PKS1830, and includes a measurement of the $^{35}$Cl/$^{37}$Cl isotopic ratio at $z$=0.89. Forthcoming publications will deal with constraints of the variations of fundamental constants, chemistry, and isotopic ratios in the $z$=0.89 absorber toward \PKS1830, based on these ALMA Cycle~0 data.

\begin{acknowledgement}
This paper makes use of the following ALMA data: ADS/JAO.ALMA\#2011.0.00405.S. ALMA is a partnership of ESO (representing its member states), NSF (USA) and NINS (Japan), together with NRC (Canada) and NSC and ASIAA (Taiwan), in cooperation with the Republic of Chile. The Joint ALMA Observatory is operated by ESO, AUI/NRAO and NAOJ.
The financial support to Dinh-V-Trung from Vietnam’s National Foundation for Science and Technology (NAFOSTED) under contract 103.08-2010.26 is gratefully acknowledged.
\end{acknowledgement}

\appendix

\section{ALMA Cycle 0 data} \label{appendix:spec}

\begin{table*}[ht]
\caption{Lines detected at $z$=0.89 toward \PKS1830\ in the ALMA Cycle 0 survey.}
\label{tab:list-all-lines}
\begin{center} \begin{tabular}{llcccccc}
\hline
 & & Rest  & Sky Freq. & $E_{low}$ & $S_{ul}$ & Dipole & Q(5.14\,K) \\
Species & Line     &Freq. & $z$=0.88582 & (K) &     & Moment   & \\
&      &(GHz) & (GHz) & &     &  (Debye)  & \\
\hline
Hydrogen cyanide & HC$^{15}$N (2-1) & 172.108 & 91.264 & 4.1 & 2.0 & 2.99 & 2.8 \\
Methanimine & CH$_2$NH (2$_{11}$-2$_{02}$) & 172.267 & 91.349 & 9.2 & 7.4 & 1.53 & 14.6 \\
Hydrogen cyanide & H$^{13}$CN (2-1) & 172.678 & 91.567 & 4.1 & 2.0 & 2.99 & 2.8 \\
Formyl radical & HCO (2$_{02}$-1$_{01}$) (hfs) & 173.377 & 91.937 & 4.2 & 2.8 & 1.36 & 11.3 \\
Formylium & H$^{13}$CO$^+$ (2-1) & 173.507 & 92.006 &  4.2 & 2.0 & 3.90 &    2.8 \\
Silicon monoxide & SiO (4-3) & 173.688 & 92.102 & 12.5 & 4.0 & 3.10 & 5.3 \\
Formylium  & HC$^{17}$O$^+$ (2-1) & 174.113 & 92.328 &  4.2 & 2.0 & 3.90 & 2.8 \\
Hydrogen isocyanide & HN$^{13}$C (2-1) & 174.179 & 92.363 & 4.2 & 2.0 & 3.05 & 2.8 \\
Ethynyl & C$_2$H (N=2-1, J=5/2-3/2) (hfs) & 174.663 & 92.619 & 4.2 & 2.8 & 0.77 & 11.3 \\
Hydrogen cyanide & HCN (2-1) & 177.261 & 93.997 & 4.3 & 2.0 & 2.99 & 2.8 \\
Hydrogen isocyanide & H$^{15}$NC (2-1) & 177.729 & 94.245 & 4.3 & 2.0 & 3.05 & 2.8 \\
Formylium & HCO$^+$ (2-1) & 178.375 & 94.588 & 4.3 & 2.0 & 3.90 & 2.8 \\
Sulfur monoxide & SO (5$_4$-4$_3$) & 178.605 & 94.710 & 15.9 & 4.9 & 1.54 & 7.6 \\
Hydroxymethylidyne & HOC$^+$ (2-1) & 178.972 & 94.904 & 4.3 & 2.0 & 2.77 & 2.8 \\
Carbon monosulfide & CS (4-3) & 195.954 & 103.909 & 14.1 & 4.0 & 1.96 & 4.7 \\
Carbon monoxide & CO (4-3) & 461.041 & 244.478 & 33.2 & 4.0 & 0.11 & 5.8 \\
Amidogen & NH$_2$ (1$_{10}$-1$_{01}$) (hfs, para) & 462.434 & 245.216 & 0.0 & 2.1 & 1.82 & 18. \\
Chloronium & H$_2^{37}$Cl$^+$ (1$_{11}$-0$_{00}$) (hfs, para) & 484.232 & 256.775 &  0.0 & 2.0 & 1.89 & 4.1 \\
Chloronium & H$_2^{35}$Cl$^+$ (1$_{11}$-0$_{00}$) (hfs, para) & 485.418 & 257.404 &  0.0&  2.0 & 1.89 & 4.1 \\
Carbon & C\,I $^3$P$_1$-$^3$P$_0$ & 492.161 & 260.980 & 0.0 & -- & -- & -- \\
Methanol & CH$_3$OH (4$_{14}$-3$_{03}$) & 492.279 & 261.042 & 13.9 & 2.4 & 1.44 & 7.0 \\
Methylidyne & CH (N=1, J=3/2-1/2) (hfs) & 532.7/536.8 & 282.5/284.6 &  0.2 & 0.83 & 1.46 &    8.1 \\
Carbon monoxide & $^{13}$CO (5-4) & 550.926 &292.142& 52.9 & 5.0 & 0.11 & 6.0 \\
Water & H$_2^{17}$O (1$_{10}$-1$_{01}$) (ortho) & 552.021 & 292.722 & 0.0 & 4.5 & 1.85 & 9.0 \\
Water & H$_2$O (1$_{10}$-1$_{01}$) (ortho) & 556.936 & 295.328 & 0.0 & 4.5 & 1.85 & 9.0 \\
Ammonia & NH$_3$ (1$_0$-0$_0$) (hfs, ortho) & 572.498 & 303.580 &  0.0 & 2.7 & 1.47 &   39.2 \\
Carbon monoxide & CO (5-4) & 576.268 & 305.580 & 55.3 & 5.0 & 0.11 & 5.8 \\
\hline
\end{tabular} 
\tablefoot{$E_{low}$ is the lower-level energy of the transition. $S_{ul}$ is the line strength. Q(5.14\,K) is the value of the partition function at a temperature of 5.14\,K.
For species with hyperfine structure (indicated by hfs), the values are given for the hyperfine component with the maximum relative intensity.}
\end{center} \end{table*}

\clearpage

\begin{figure*}[h] \begin{center}
\includegraphics[width=\textwidth]{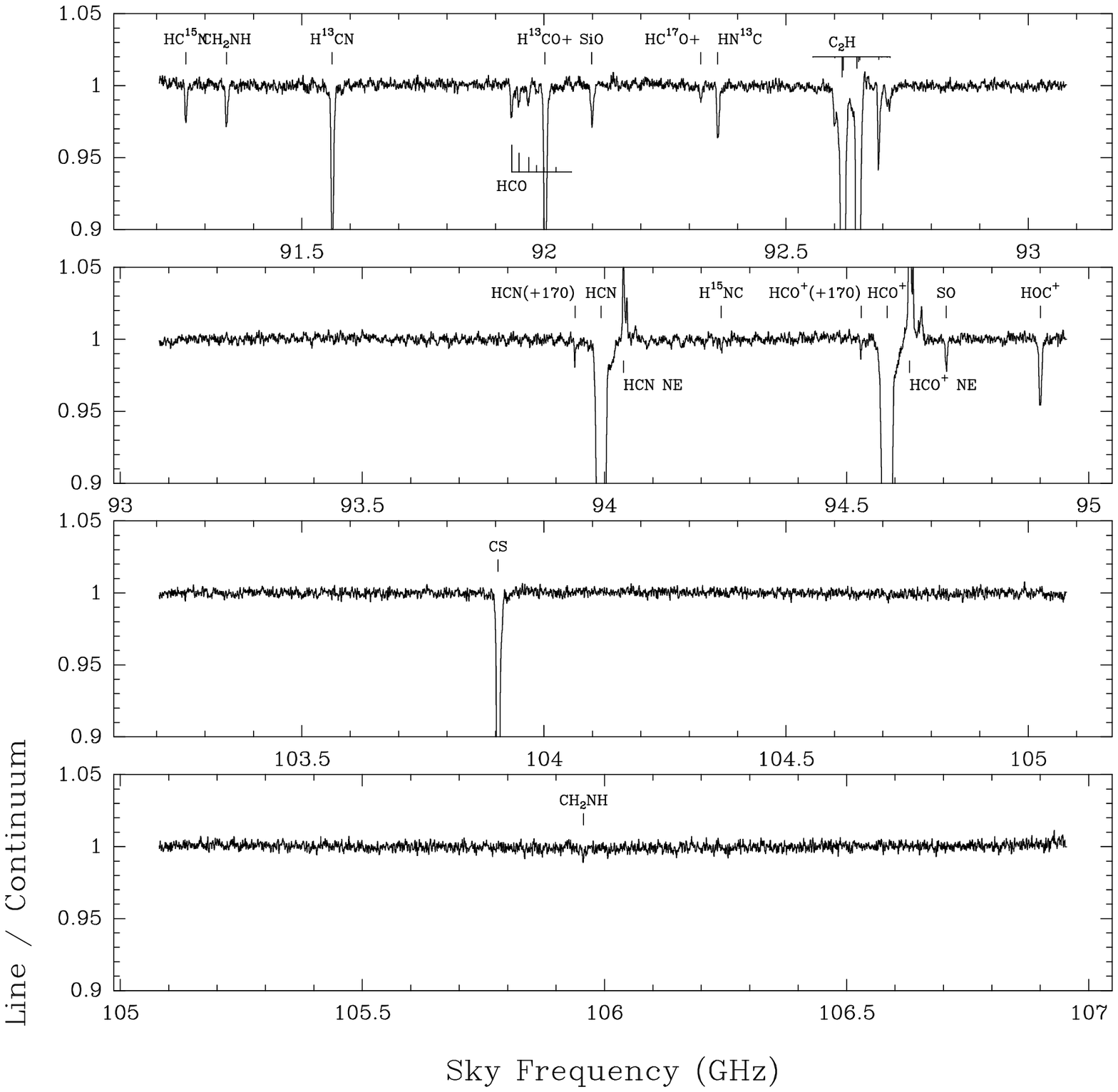}
\caption{ALMA Cycle 0 B3--100\,GHz spectra, after self-bandpass of the SW spectra by the NE spectra. Absorption lines from the SW line-of-sight appear as $<$1-feature, while the absorption lines from the NE line-of-sight appear as $>$1-feature.}
\label{fig:LO100}
\end{center} \end{figure*}

\begin{figure*}[h] \begin{center}
\includegraphics[width=\textwidth]{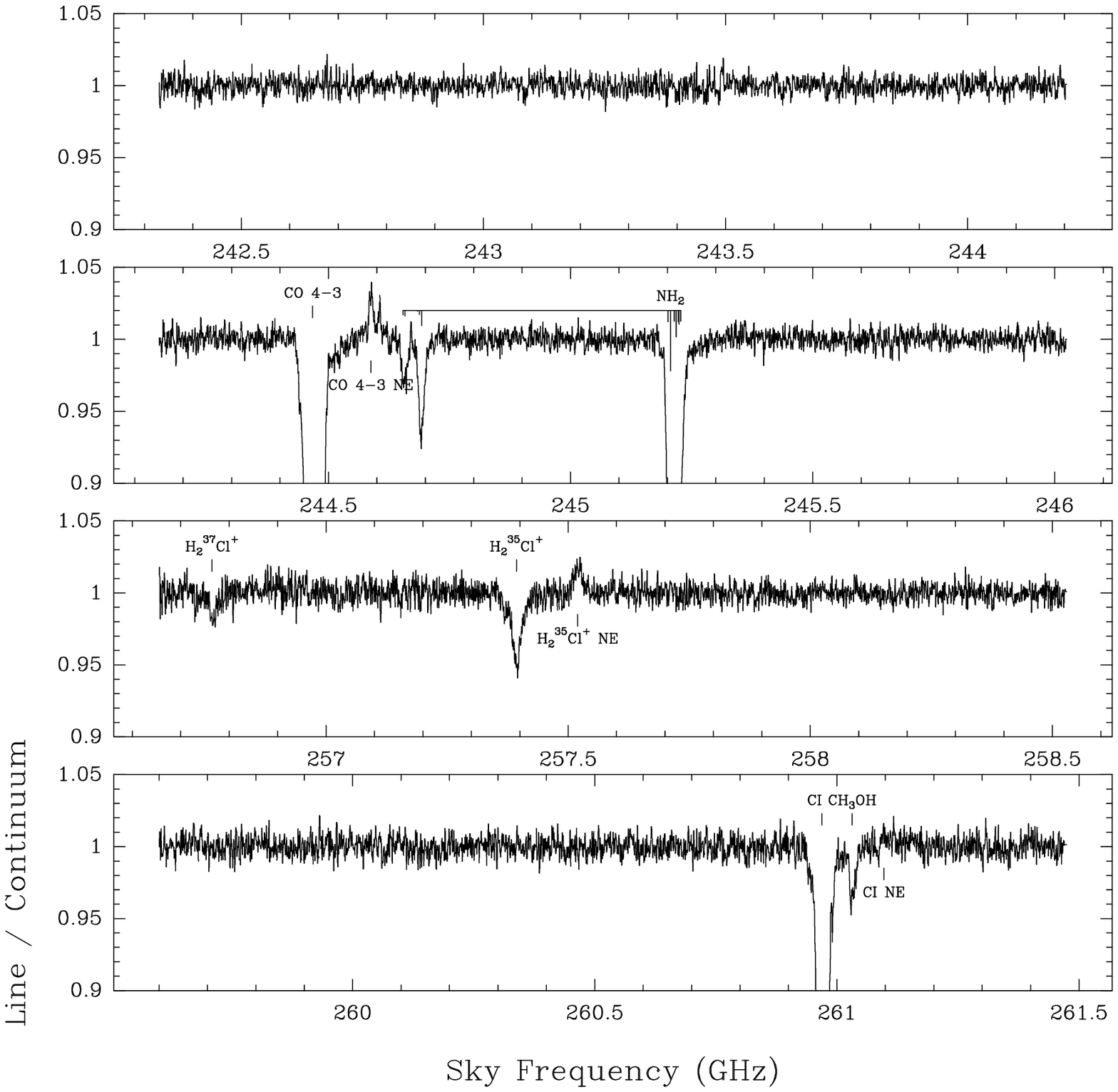}
\caption{ALMA Cycle 0 B6--250\,GHz spectra, after self-bandpass of the SW spectra by the NE spectra. Absorption lines from the SW line-of-sight appear as $<$1-feature, while the absorption lines from the NE line-of-sight appear as $>$1-feature.}
\label{fig:LO250}
\end{center} \end{figure*}

\begin{figure*}[h] \begin{center}
\includegraphics[width=\textwidth]{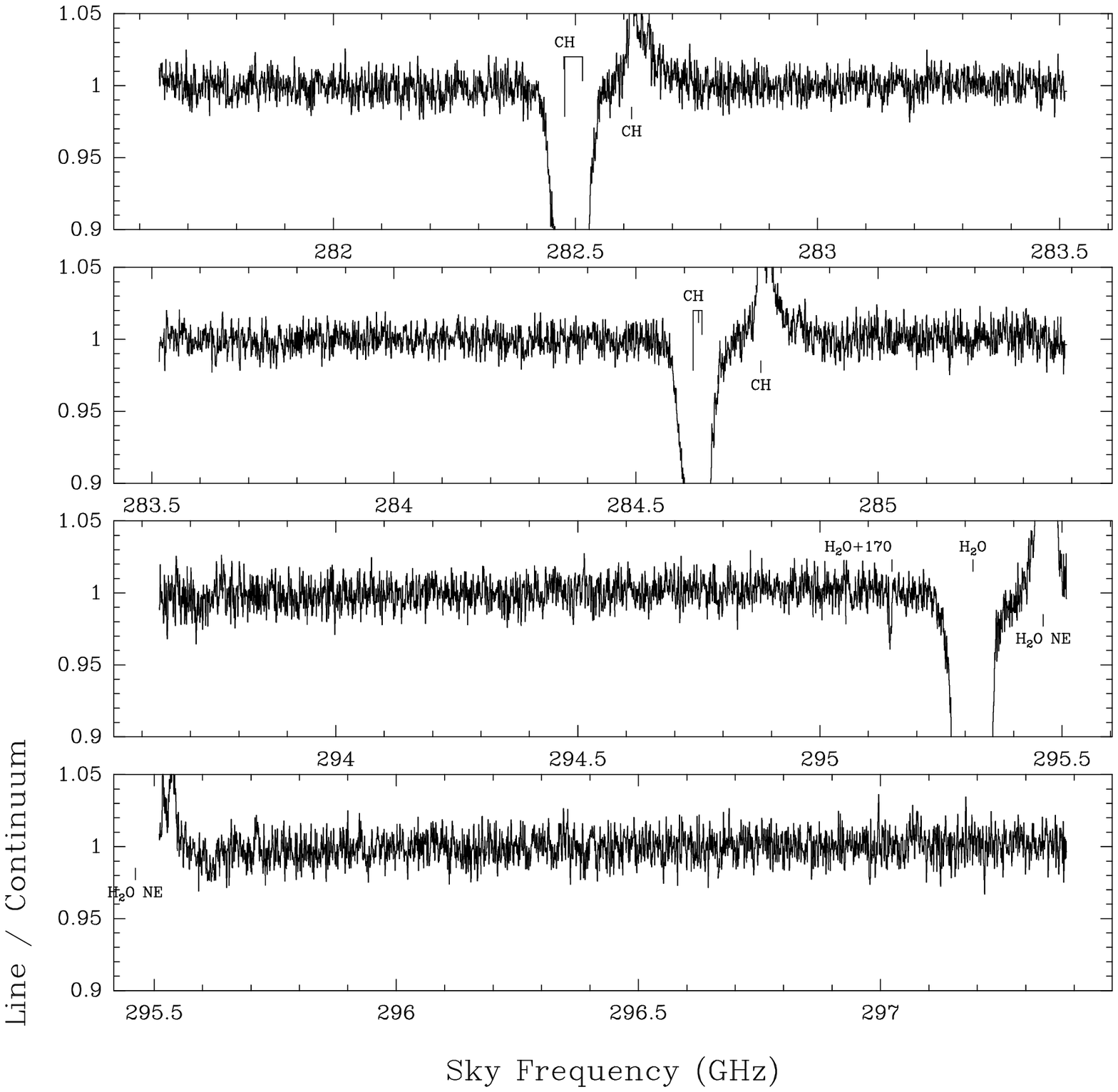}
\caption{ALMA Cycle 0 B7--290\,GHz spectra, after self-bandpass of the SW spectra by the NE spectra. Absorption lines from the SW line-of-sight appear as $<$1-feature, while the absorption lines from the NE line-of-sight appear as $>$1-feature.}
\label{fig:LO290}
\end{center} \end{figure*}

\begin{figure*}[h] \begin{center}
\includegraphics[width=\textwidth]{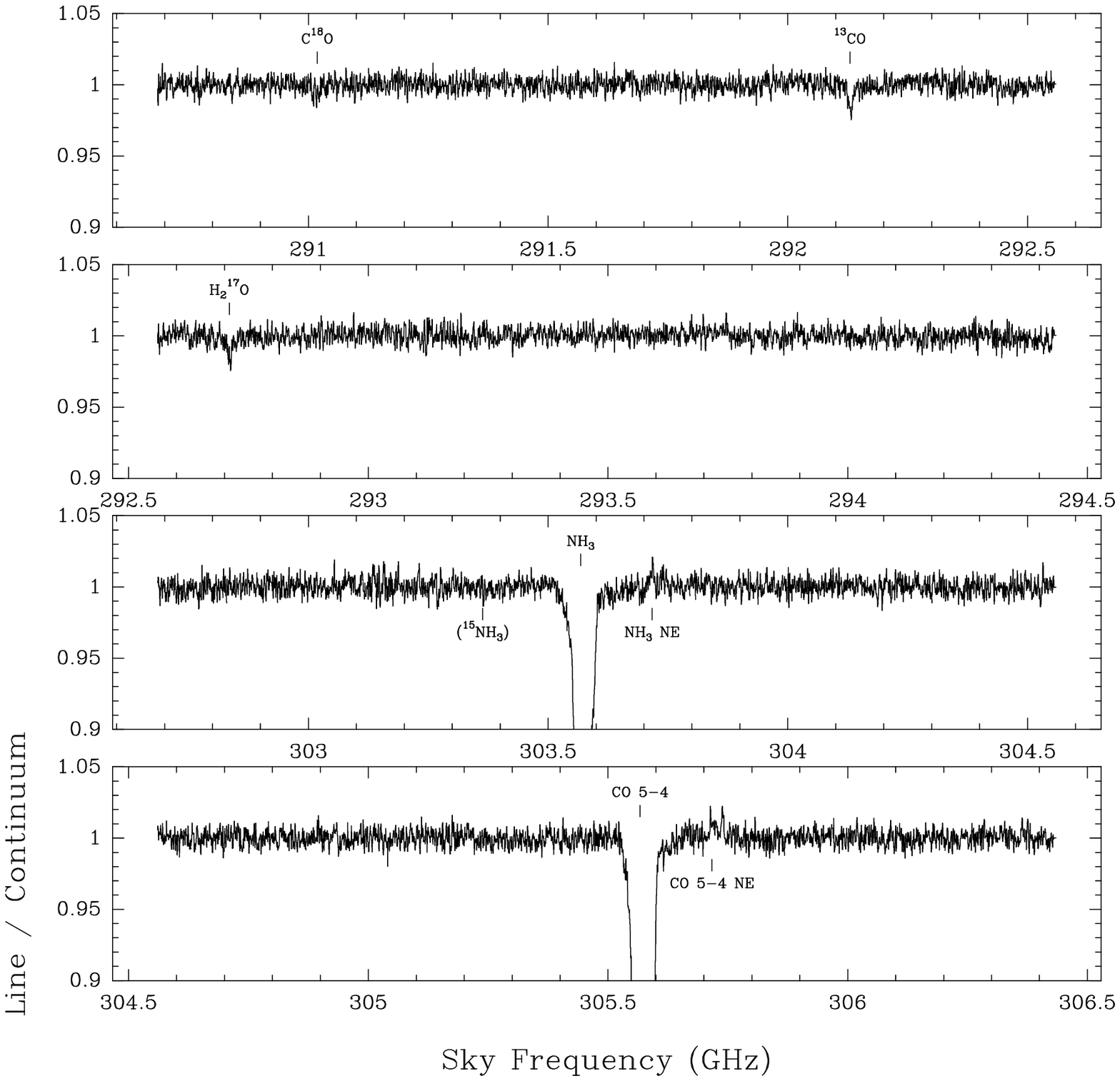}
\caption{ALMA Cycle 0 B7--300\,GHz spectra, after self-bandpass of the SW spectra by the NE spectra. Absorption lines from the SW line-of-sight appear as $<$1-feature, while the absorption lines from the NE line-of-sight appear as $>$1-feature.}
\label{fig:LO300}
\end{center} \end{figure*}

\clearpage

\begin{table}[ht]
\caption{O$_3$ atmospheric lines in the observed bands.} \label{tab:O3atmline}
\begin{center} \begin{tabular}{crc}
\hline
Tuning & \multicolumn{1}{c}{Rest freq.} &  Blend with \\
       & \multicolumn{1}{c}{(GHz)}      & a  $z$=0.89 line \\
\hline
B3--100\,GHz &  93.844 & no \\
      &  93.955 & no \\
      & 103.878 & no \\
B6--250\,GHz & 242.319 & no \\
      & 243.454 & no \\
      & 244.158 & no \\
      & 258.202 & no \\
B7--290\,GHz & 282.837 & no \\
B7--300\,GHz & 290.975 & no \\
      & 293.548 & no \\
      & 293.171 & no \\
      & 303.165 & no \\
\hline
\end{tabular}
\tablefoot{The {\em a posteriori} Doppler correction (Earth motion relative to \PKS1830 direction) makes the atmospheric lines shift slightly (few MHz) from their rest frequencies, depending on the epoch.}
\end{center} \end{table}

\begin{figure}[h] \begin{center}
\includegraphics[width=8.8cm]{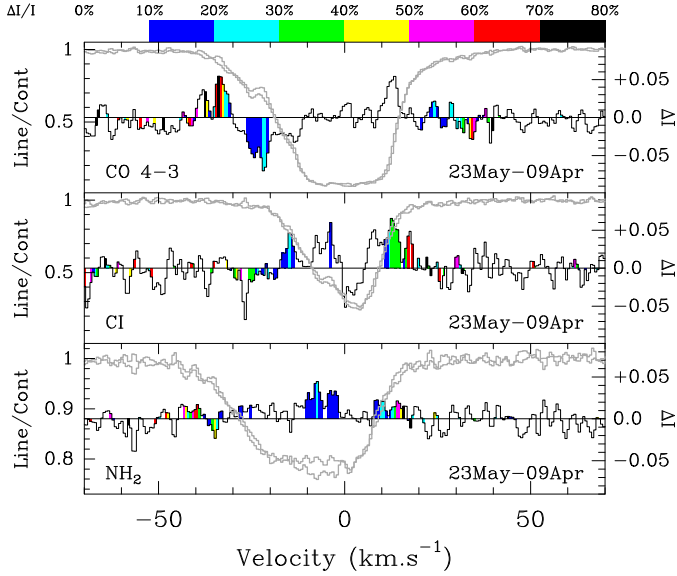}
\caption{Time variations toward the SW image between 2012 April 09 and 2012 May 23.}
\label{fig:timvar-23May09Apr}
\end{center} \end{figure}

\begin{figure}[h] \begin{center}
\includegraphics[width=8.8cm]{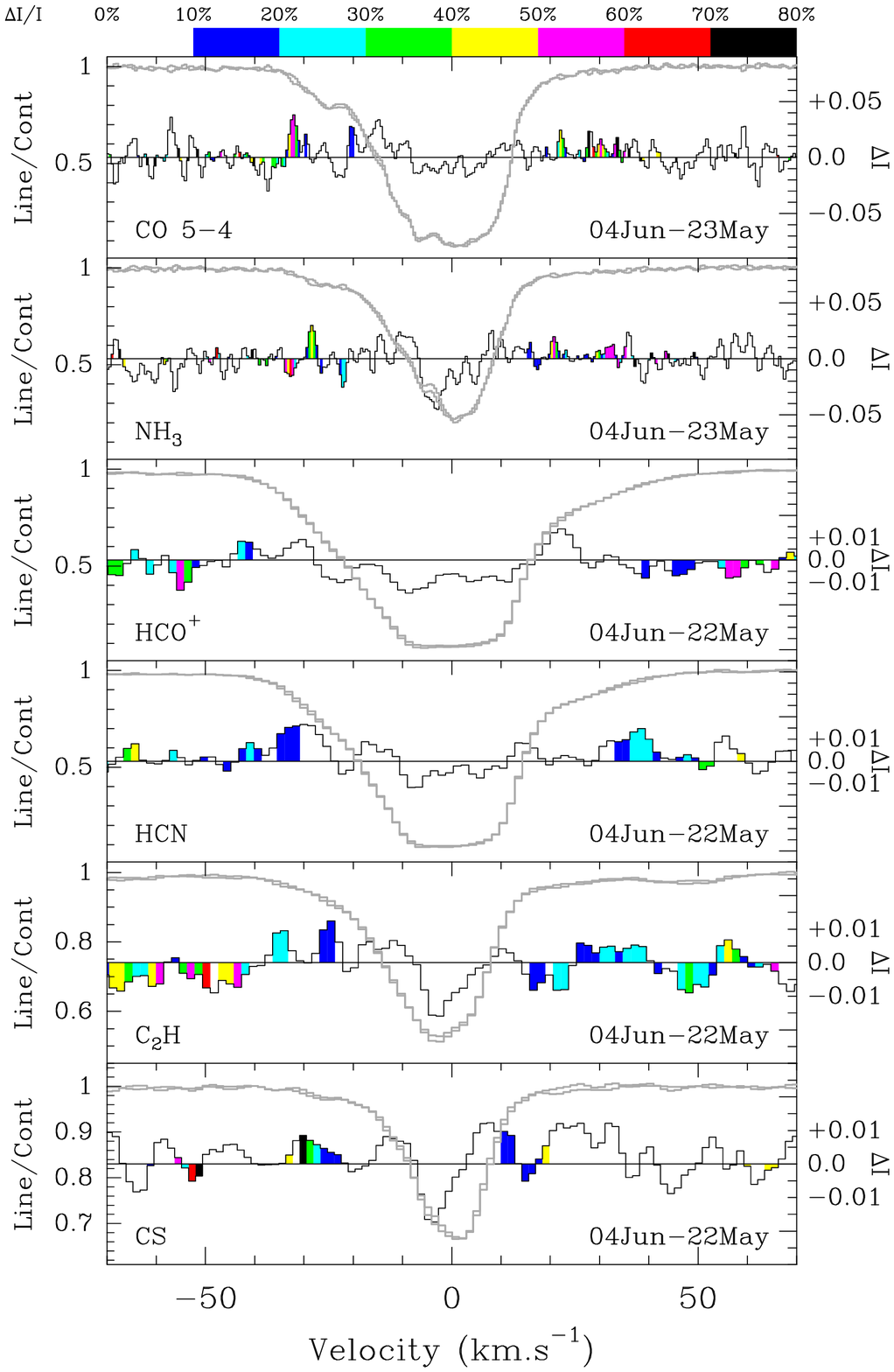}
\caption{Time variations toward the SW image between 2012 May 22 and 2012 Jun 04.}
\label{fig:timvar-04Jun22May}
\end{center} \end{figure}

\begin{figure}[h] \begin{center}
\includegraphics[width=8.8cm]{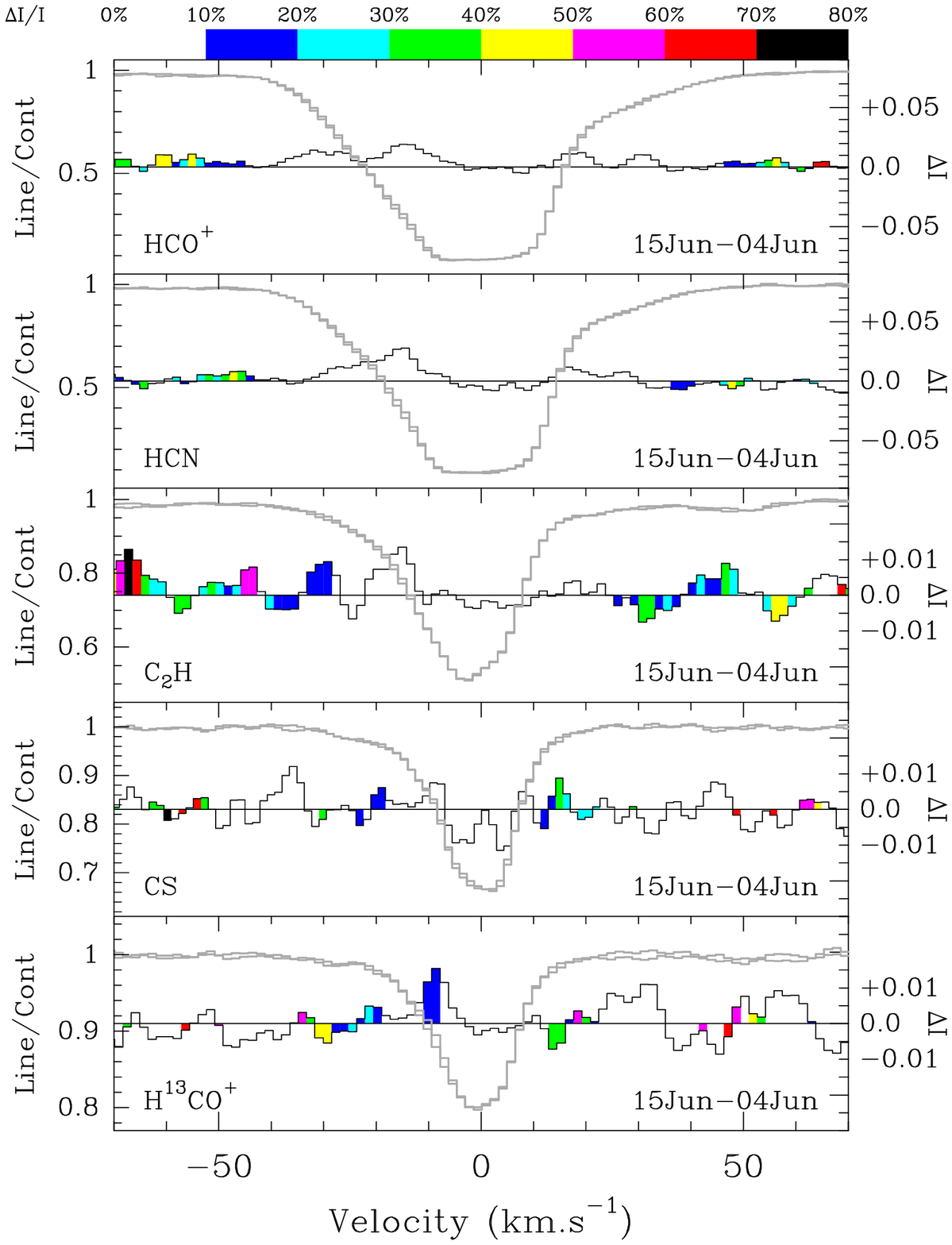}
\caption{Time variations toward the SW image between 2012 Jun 04 and 2012 Jun 15.}
\label{fig:timvar-15Jun04Jun}
\end{center} \end{figure}

\begin{figure}[h] \begin{center}
\includegraphics[width=8.8cm]{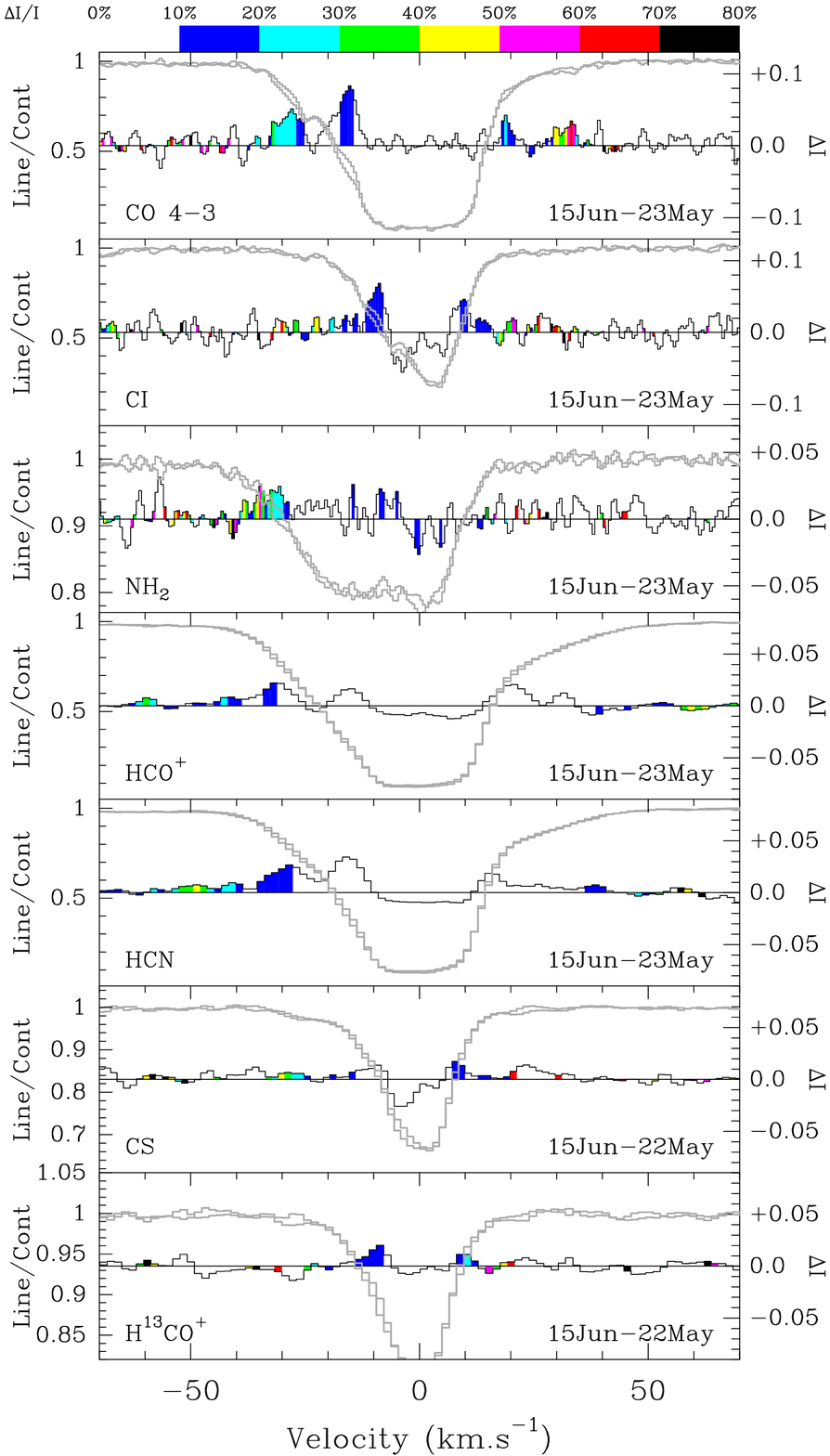}
\caption{Time variations toward the SW image between 2012 May 22 and 2012 Jun 15.}
\label{fig:timvar-15Jun22May}
\end{center} \end{figure}

\begin{figure*}[h] \begin{center}
\includegraphics[width=\textwidth]{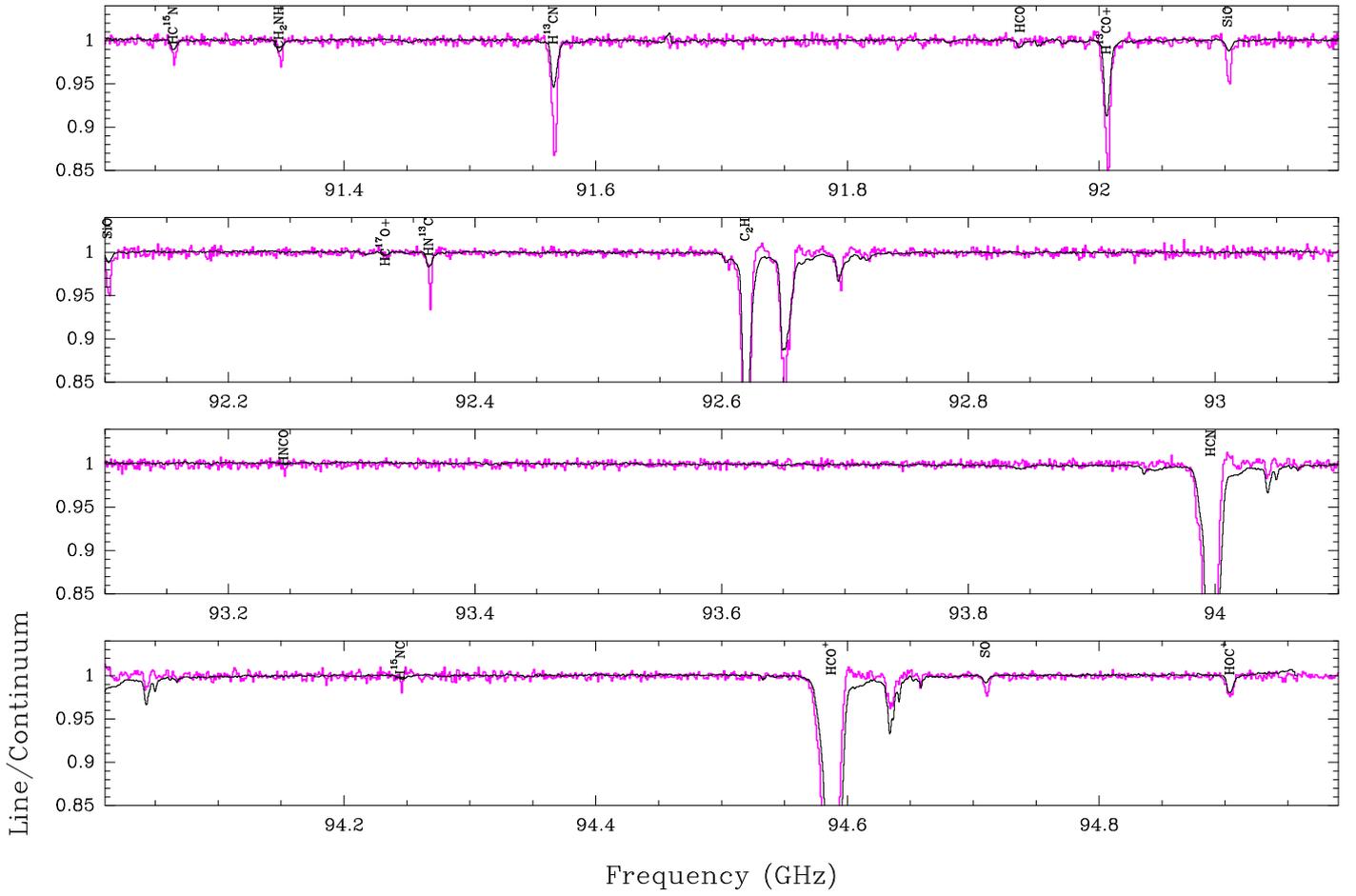}
\caption{Composite ALMA B3 spectra (observed in 2012, in {\em black}) overlaid on top of the ATCA-3mm spectra obtained in July 2011 (in {\em magenta}) by \cite{mul13}. The ALMA spectra has been recomposed using the spectra from both images (unresolved in the ATCA observations), assuming $\Re$=$F_{NE}$/$F_{SW}$=1.35, $f_c$(SW)=92\%, and $f_c$(NE)=100\%.}
\label{fig:overlayATCA}
\end{center} \end{figure*}

\begin{figure}[h] \begin{center}
\includegraphics[width=8.cm]{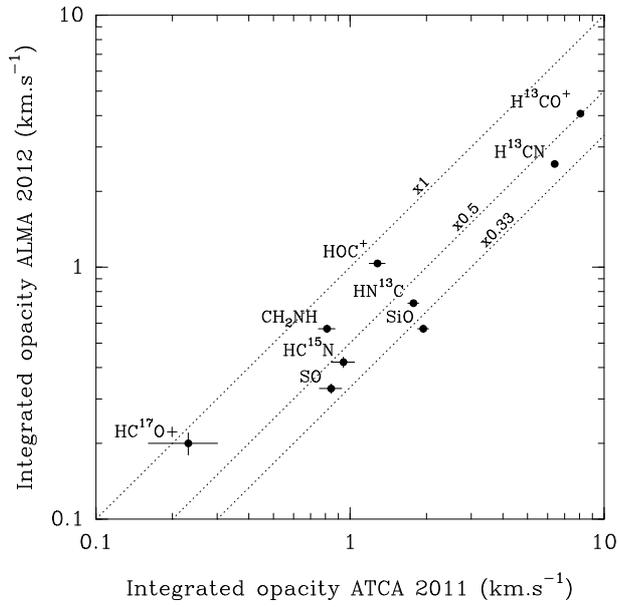}
\caption{Comparison of the integrated opacities of the same lines between ATCA observations in July 2011 and ALMA observations in spring 2012.}
\label{fig:timevar-comparo}
\end{center} \end{figure}

\end{document}